\documentclass[aps,pra,twocolumn,groupedaddress,nofootinbib,superscriptaddress]{revtex4-1}

\usepackage{amsmath,mleftright,mathtools}
\usepackage{bm}
\usepackage{stix}
\usepackage{graphicx}
\usepackage{xspace}
\usepackage{units}
\usepackage[colorlinks,linkcolor=blue,citecolor=blue,urlcolor=blue]{hyperref}
\usepackage[dvipsnames]{xcolor}
\usepackage{cleveref}
\usepackage{array}   
\newcolumntype{L}{>{$}l<{$}} 
\usepackage{dcolumn}
\newcolumntype{d}[1]{D{.}{.}{#1}}

\def\bra{\langle}
\def\ket{\rangle}
\def\a{\alpha}
\def\b{\beta}
\def\d{\delta}
\def\e{\epsilon}
\def\l{\lambda}
\def\m{\mu}
\def\n{\nu}
\def\t{\tau}
\def\ua{\uparrow}
\def\da{\downarrow}

\newcommand{\bt}{\bar{t}}

\newcommand{\s}{\sigma}
\graphicspath{{./Figs/}}
\mathchardef\mhyphen="2D \newcommand{\nn}{\nonumber}

\newcommand{\be}{\begin{equation}}
\newcommand{\ee}{\end{equation}}

\newcommand{\ii}{\mathrm{i}}

\newcommand{\mv}{\mathbfit{v}}
\newcommand{\mw}{\mathbfit{w}}
\newcommand{\mh}{\mathbfit{h}}

\newcommand{\mPsi}{\bm{\mathit{\Psi}}}

\newcommand{\mPi}{\mathbfit{\Pi}}
\newcommand{\cG}{\mathcal{G}}

\newcommand{\mcG}{\bm{\mathcal{G}}}

\newcommand{\mP}{\mathbfit{P}}
\newcommand{\mrho}{\bm{\mathit{\rho}}}
\newcommand{\mchi}{\bm{\mathit{\chi}}}

\newcommand{\md}{\bm{\delta}}

\newcommand{\vac}{\phi}

\begin{document}
\title{Three-particle ladder  method for photoinduced dynamics of organic molecules}
\title{Photoinduced dynamics of organic molecules using 
nonequilibrium Green's functions with second-Born, $GW$, $T$-matrix and three-particle ladder 
correlations}
\author{Y. Pavlyukh}
\affiliation{Dipartimento di Fisica, Universit{\`a} di Roma Tor Vergata, Via della Ricerca Scientifica 1,
00133 Rome, Italy}
\author{E. Perfetto}
\affiliation{Dipartimento di Fisica, Universit{\`a} di Roma Tor Vergata, Via della Ricerca Scientifica 1,
00133 Rome, Italy}
\affiliation{INFN, Sezione di Roma Tor Vergata, Via della Ricerca Scientifica 1, 00133 Rome, Italy}
\author{G. Stefanucci}
\affiliation{Dipartimento di Fisica, Universit{\`a} di Roma Tor Vergata, Via della Ricerca Scientifica 1,
00133 Rome, Italy}
\affiliation{INFN, Sezione di Roma Tor Vergata, Via della Ricerca Scientifica 1, 00133 Rome, Italy}
\date{\today}
\begin{abstract}  
The ultrafast hole dynamics triggered by the photoexcitation of molecular targets is a
highly correlated process even for those systems, like organic molecules, having a weakly
correlated ground state. We here provide a unifying framework and a numerically efficient
matrix formulation of state-of-the-art non-equilibrium Green's function (NEGF) methods
like second-Born as well as $GW$ and $T$-matrix without and {\em with} exchange
diagrams. Numerical simulations are presented for a paradigmatic, exactly solvable
molecular system and the shortcomings of the established NEGF methods are highlighted.  We
then develop a NEGF scheme based on the Faddeev treatment of three-particle correlations;
the exceptional improvement over established methods is explained and demonstrated. The
Faddeev NEGF scheme scales linearly with the maximum propagation time, thereby opening
prospects for femtosecond simulations of large molecules.
\end{abstract}
\maketitle
\section{Introduction}                            
The study of nonequilibrium phenomena in correlated materials has recently become one of
the most active and exciting branches of atomic, molecular and condensed matter
physics. This is largely due to advances in light sources and time-resolved spectroscopies
on the ultrashort time scales~\cite{zhang_dynamics_2014}, which made it possible not only
to observe and describe but also to design systems with new remarkable properties by
coupling them to external electromagnetic fields~\cite{lepine_attosecond_2014}. In a long
time perspective they may lead to practical applications having a huge societal 
impact~\cite{kraus_ultrafast_2018}.

As an example, consider the quantum evolution of an organic molecule initially in its
weakly correlated ground-state and then perturbed by an ultra-short (sub-fs) weak XUV
pulse~\cite{calegari_ultrafast_2014,iablonskyi_observation_2017,lara-astiaso_attosecond_2018,herve_ultrafast_2020}.
The target molecule undergoes a transition to an excited one-hole state through the
emission of a single electron. The resulting cationic state can no longer be characterized
as weakly correlated. In fact, immediately after the excitation quantum scattering
processes mediated by the Coulomb interaction start to roll. They promote the decay of the
left behind hole into a two-hole and one-particle ($2h$-$1p$) state. Thus, in contrast to
the initial non-degenerate ground state, the system is now in a superposition of a large
number of \emph{quasi-degenerate states} whose energies and mutual interactions represent
a formidable challenge for the theory.  In particular, this is true for methods based on
density functional theory as they rely on ground state correlations only. It is also a
challenge for wave-function methods. They can, in principle, deal with
multi-configurational ionized states (static
correlations)~\cite{kuleff_multielectron_2005,szalay_multiconfiguration_2011,popova-gorelova_imaging_2016}
and systematically treat $2h$-$1p$, $3h$-$2p$, etc. configurations, thus making the
approach accurate and predictive for small molecular systems. However, the inclusion of
dynamical correlations (quasi-particle dressing in physics therminlogy) remains a
difficult numerical task.

Other challenges for the theory include the treatment of large molecular systems, where
nuclear and collective electronic excitations emerge as important scattering
channels~\cite{schuler_electron_2016,usenko_femtosecond_2016}, as well as the description
of processes with a variable number of particles as in transport~\cite{me-book1,me-book2}
and photoemission
experiments~\cite{cardona_photoemission_1978,freericks_theoretical_2009,pavlyukh_single-_2015}.
Methods that can deal with all theses ingredients on equal footing are still in their
infancy; developments in the realm of wave-function
expansions~\cite{ruberti_b-spline_2014,ruberti_full_2018,ruberti_multi-channel_2018,pathak_time-dependent_2020}
and time-dependent DFT are certainly
foreseeable~\cite{andreussi_carotenoids_2015,nisoli_attosecond_2017}.

The nonequilibrium Green's function (NEGF)
theory~\cite{stefanucci_nonequilibrium_2013,balzer_nonequilibrium_2013} is another fertile
playground for the development of efficient methods.  Its main variable, namely the
single-particle Green's function, naturally appears in the observables characterizing the
aforementioned phenomena, and the inclusion of static and dynamical electronic
correlations as well as interactions with other quasiparticles of bosonic nature, such as
plasmons and vibrational modes, is possible through the exact resummation of diagrammatic
expansions to infinite order in the interactions strength.

The NEGF versatility, however, comes at the cost of dealing with two-times
correlators. The time-evolution of any quantum systems is described by the so called
Kadanoff-Baym equations
(KBE)~\cite{stefanucci_nonequilibrium_2013,balzer_nonequilibrium_2013} for the Green's
function. The KBE are nonlinear first-order integro-differential equations scaling
cubically with the physical propagation time, thereby making it difficult to resolve small
energy scales associated with phonons, magnons, etc. A less severe quadratic scaling can
be achieved by means of the so-called Generalized Kadanoff-Baym Ansatz
(GKBA)~\cite{lipavsky_generalized_1986} which allows for reducing the KBE to a single
equation of motion for the one-particle density
matrix~\cite{karlsson_generalized_2018}. Recent applications of the NEGF + GKBA approach
include the nonequilibrium dynamics~\cite{hermanns_hubbard_2014,schlunzen_dynamics_2016}
and many-body localization~\cite{bar_lev_dynamics_2014} of Hubbard clusters,
time-dependent quantum transport~\cite{latini_charge_2014,cosco2020spectral,Riku2021},
real-time description of the Auger decay~\cite{covito_real-time_2018}, excitonic
insulators out of equilibrium~\cite{tuovinen_comparing_2020}, equilibrium absorption of
sodium clusters~\cite{pal_optical_2011}, transient
absorption~\cite{perfetto_first-principles_2015,perfetto_nonequilibrium_2015,sangalli_nonequilibrium_2016,pogna_photo-induced_2016}
and carrier dynamics~\cite{sangalli_ultra-fast_2015,perfetto_first-principles_2016} of
semiconductors. A tremendous progress has been recently achieved in further reducing the
scaling to the ideal linear law and establishing that the method is applicable for
state-of-the-art diagrammatic approximations like the second-Born (2B), $GW$ and
$T$-matrix (both in the $ph$ and $pp$ channels)~\cite{joost_g1-g2_2020}.  These
approximations have been extensively tested in the past for model and realistic systems in
the neutral state, both by solving full Kadanoff-Baym
equations~\cite{dahlen_solving_2007,myohanen_many-body_2008,myohanen_kadanoff-baym_2009,puig_von_friesen_kadanoff-baym_2010,
  friesen_artificial_2010,sakkinen_kadanoffbaym_2012} and by using
GKBA~\cite{joost_g1-g2_2020,tuovinen_comparing_2020,murakami_ultrafast_2020}.  However,
they loose accuracy in the description of photoionization-induced dynamics even for
systems having a weakly correlated ground state.

Let us return to our initial picture of the $1h\rightarrow 2h$-$1p$ scattering in
photoexcited molecular targets. In a realistic scenario one has to deal with recurrent
scatterings of this kind. Mathematically this is treated by the resummation of certain
classes of Feynman diagrams. One may focus on the fate of one particle and one hole in the
final state and disregard other interactions, schematically indicated as $h\rightarrow
(p+h)+h$. Depending on which hole $h$ is paired with the particle $p$ in the final state
we end up with either the $GW$ approximation or the $T$-matrix approximation in the $ph$
channel (henceforth $T^{ph}$).  Alternatively, one may elect to describe the interactions
between two-holes (or particles) in the final channel, schematically indicated as
$h\rightarrow (h+h)+p$, leading to the so-called $T$-matrix approximation in the $pp$
channel (henceforth $T^{pp}$)~\cite{pavlyukh_initial_2013}.  All these approximations
treat either a hole or a particle as spectator, i.\,e., they ignore three-particle
correlations.  Such limitation has a profound impact in the description of fundamental
physical processes. This is especially true in the presence of (near) degeneracies between
the involved electronic states.  In the case of the inter-valence hole migration the
quasi-degeneracies are due to spin degrees of freedom. The multitude of spin-states in the
$1h\rightarrow 2h$-$1p$ scattering scenario is not accounted for by the conventional $GW$
and $T$-matrices approximations.

In this work we apply all conventional approximations to study the inner-valence hole
migration in the glycine molecule.  The numerical simulations clearly show that none of
these methods is capable to describe the quantum beating associated with transitions
between different $2h$-$1p$ states.  A resolution within NEGF is achievable by explicitly
correlating the three-particle states. The so called three-particle ladder approximation
has been first explored in the context of nuclear physics~\cite{barbieri_faddeev_2001} and
it leads to the well known Faddeev
equations~\cite{faddeev_scattering_1961,ethofer_six-point_1969}.  These equations have
been applied to model~\cite{potthoff_three-particle_1994},
atomic~\cite{barbieri_quasiparticles_2007} and small molecular
systems~\cite{degroote_faddeev_2011}. However, to the best of our knowledge, the Faddeev
equations have never been investigated in the context of the NEGF formalism.

The main achievement of our work is the development of a NEGF+GKBA method based on the
three-particle ladder diagrams.  For the purpose of a self-contained exposition we first
introduce the 2B, $GW$ and $T$-matrix approximations. In Sec.~\ref{sec:derivation} we
provide a simple and concise derivation of the equations of motion, cast the equations in
a numerically efficient matrix form and highlight the common underlying mathematical
structure of all these approximations.  In Sec.~\ref{sec:2h-1p} we present the
full-fledged three-particle method, henceforth refered to as the Faddeev approximation.
Its derivation relies on the extension of the GKBA to high-order Green's functions.
Conventional and Faddeev approximations are benchmarked against the exact photoinduced
electron dynamics in the paradigmatic glycine molecule, finding an excellent agreement for
the latter, see Section~\ref{sec:gly:III}.  Noteworthy, the numerical solution of the
Faddeev-GKBA method scales linearly with the maximum propagation time.  In
Sec.~\ref{sec:conclusions} we recapitulate our finding and propose systems and
experimental scenarios where the method is particularly relevant.

\section{Unifying formulation of the GKBA equations for state-of-the-art methods}%
  \label{sec:derivation}
\begin{table*}[t]
  \caption{\label{tab:1} Definitions of the two-particle 2-rank 
  tensors. The vertically grouped indices are combined into one
    super-index. Here $h\equiv h_\text{HF}$ for brevity.}
  \renewcommand{\arraystretch}{1.4}
  \begin{ruledtabular}
    \begin{tabular}{LLLLL}
      \text{Quantity}& \text{2B}& GW\; +(X) & T^{pp}\;+(X) & 
      T^{ph}\;+(X)\\\hline
      \mcG & 
      \mcG_{\begin{subarray}{c}13\\24\end{subarray}}=\mathcal{G}_{4132} 
      &\mcG_{\begin{subarray}{c}13\\24\end{subarray}}=\mathcal{G}_{4132}
      &\mcG_{\begin{subarray}{c}13\\24\end{subarray}}=\mathcal{G}_{1234} 
      &\mcG_{\begin{subarray}{c}13\\24\end{subarray}}=\mathcal{G}_{1432}\\  [6pt] 
      \mh
      &\mh_{\begin{subarray}{c}13\\24\end{subarray}}=h_{13}\delta_{42}-\delta_{13}h_{42}
      &\mh_{\begin{subarray}{c}13\\24\end{subarray}}=h_{13}\delta_{42}-\delta_{13}h_{42}
      &\mh_{\begin{subarray}{c}13\\24\end{subarray}}=h_{13}\delta_{24}+\delta_{13}h_{24}
      &\mh_{\begin{subarray}{c}13\\24\end{subarray}}=h_{13}\delta_{42}-\delta_{13}h_{42}\\ 
       \mv & \mv_{\begin{subarray}{c}13\\24\end{subarray}}=v_{1432} 
      &\mv_{\begin{subarray}{c}13\\24\end{subarray}}=v_{1432} 
      &\mv_{\begin{subarray}{c}13\\24\end{subarray}}=v_{1243}
      &\mv_{\begin{subarray}{c}13\\24\end{subarray}}=v_{1423}\\ [6pt]
      \mw & 
      \mw_{\begin{subarray}{c}13\\24\end{subarray}}=v_{1432}-v_{1423} 
      &\mw_{\begin{subarray}{c}13\\24\end{subarray}}=v_{1432}-(v_{1423}) 
      &\mw_{\begin{subarray}{c}13\\24\end{subarray}}=v_{1243}-(v_{1234})
      &\mw_{\begin{subarray}{c}13\\24\end{subarray}}=v_{1423}-(v_{1432})\\[6pt]
      \mrho^{<} 
      &\mrho^{<}_{\begin{subarray}{c}13\\24\end{subarray}}=\rho^<_{13}\rho^>_{42}
      &\mrho^{<}_{\begin{subarray}{c}13\\24\end{subarray}}=\rho^<_{13}\rho^>_{42}
      &\mrho^{<}_{\begin{subarray}{c}13\\24\end{subarray}}=\rho^<_{13}\rho^<_{24}
      &\mrho^{<}_{\begin{subarray}{c}13\\24\end{subarray}}=\rho^<_{13}\rho^>_{42}\\ 
      \mrho^{>}
      &\mrho^{>}_{\begin{subarray}{c}13\\24\end{subarray}}=\rho^>_{13}\rho^<_{42}
      &\mrho^{>}_{\begin{subarray}{c}13\\24\end{subarray}}=\rho^>_{13}\rho^<_{42}
      &\mrho^{>}_{\begin{subarray}{c}13\\24\end{subarray}}=\rho^>_{13}\rho^>_{24}
      &\mrho^{>}_{\begin{subarray}{c}13\\24\end{subarray}}=\rho^>_{13}\rho^<_{42}\\ 
      a& 0 & -1 &  1 &  1
       \end{tabular}
    \end{ruledtabular}
\end{table*}
Let us start from a generic fermionic Hamiltonian
\be
\hat H(t)=\sum_{ij} h_{ij}(t)\hat{d}_{i}^\dagger \hat{d}_{j}
+\frac12\sum_{ijmn}v_{ijmn}(t)
  \hat{d}_{i}^\dagger \hat{d}_{j}^\dagger \hat{d}_{m}\hat{d}_{n},
  \label{ham}
\ee
where $h_{ij}$ stands for the one-body part and $v_{ijmn}$ is the Coulomb interaction
tensor, they are time-dependent in general. The time-dependence in $h_{ij}(t)$ originates,
for instance, from the coupling to external fields, whereas the time-dependence in
$v_{ijmn}(t)$ could be due to the adiabatic switching protocol adopted to generate a
correlated initial state. Below, we skip the time-arguments if they are not essential for
the discussion. The indices $i,\,j$, etc. comprise a spin index and an orbital index which
(without any loss of generality) is associated with some localized basis functions, but it
is straightforward to reformutate the equations in, e.\,g., plane-wave basis or any other
suitable basis. In this work we consider a spin symmetric single-particle Hamiltonian and
a spin-independent interaction. Making explicit the spin-dependence this implies that
$h_{i\s_{1}j\s_{2}}=\d_{\s_{1}\s_{2}}h_{ij}$ and
\be
v_{i\s_{1}j\s_{2}m\s_{3}n\s_{4}}=\d_{\s_{1}\s_{4}}\d_{\s_{2}\s_{3}}
v_{ijmn}.
\label{vspin}
\ee

The lesser and the greater Green's functions (GFs) are defined as
\begin{subequations}
\begin{align}
  G^{<}_{ij}(t,t')&=+\ii \big\langle \hat{d}_j^\dagger(t')\hat{d}_i(t)\big\rangle,\\
  G^{>}_{ij}(t,t')&=-\ii \big\langle \hat{d}_i(t)\hat{d}_j^\dagger(t')\big\rangle,
\end{align}
\end{subequations}
and fulfill the symmetry relation
$G^{\lessgtr}(t_1,t_2)=-\left[G^{\lessgtr}(t_2,t_1)\right]^\dagger$.  They carry
information on the single-particle spectra and occupations.  The generalized Kadanoff-Baym
Ansatz (GKBA)~\cite{lipavsky_generalized_1986} factorizes these two independent
ingredients, see Appendix~\ref{gkbaintroapp},
\begin{align}
  G^{\lessgtr}(t_1,t_2)&=-G^{R}(t_1,t_2)\rho^{\lessgtr}(t_2)+\rho^{\lessgtr}(t_1)G^{A}(t_1,t_2)
  \label{eq:gkba}
\end{align}
so that the greater/lesser density matrices become our main single-time variables
\be
\rho_{ij}^{\lessgtr}(t)=-\ii G^{\lessgtr}_{ij}(t,t)\quad\quad
[\rho^{>}_{ij}=\rho^{<}_{ij}-\delta_{ij}].\label{eq:def:rho}
\ee

Using the GKBA the KBE are reduced to an equation of motion for the density matrix
\begin{align}
  \frac{d}{dt}\rho^<(t)&=-\ii 
  \big[h_\text{HF}(t),\rho^<(t) \big]-
  \left(I(t)+I^\dagger(t)\right)
  \label{eomrho}
\end{align}
provided that the retarded ($G^{R}$) and advanced ($G^{A}$) Green's functions are
approximated as functional of $\rho^{<}$. In this work we consider the Hartree-Fock
functional form
\begin{align}
G^{R}(t,t')=-\ii\theta(t-t')
T\left\{e^{-\ii \int_{t'}^t d\tau\, h_\text{HF}(\tau)}\right\},
\label{hfgr}
\end{align}
and hence $G^{\rm A}(t,t')=[G^{\rm R}(t',t)]^{\dag}$.
\begin{figure}[t!]
\centering  \includegraphics[width=0.99\columnwidth]{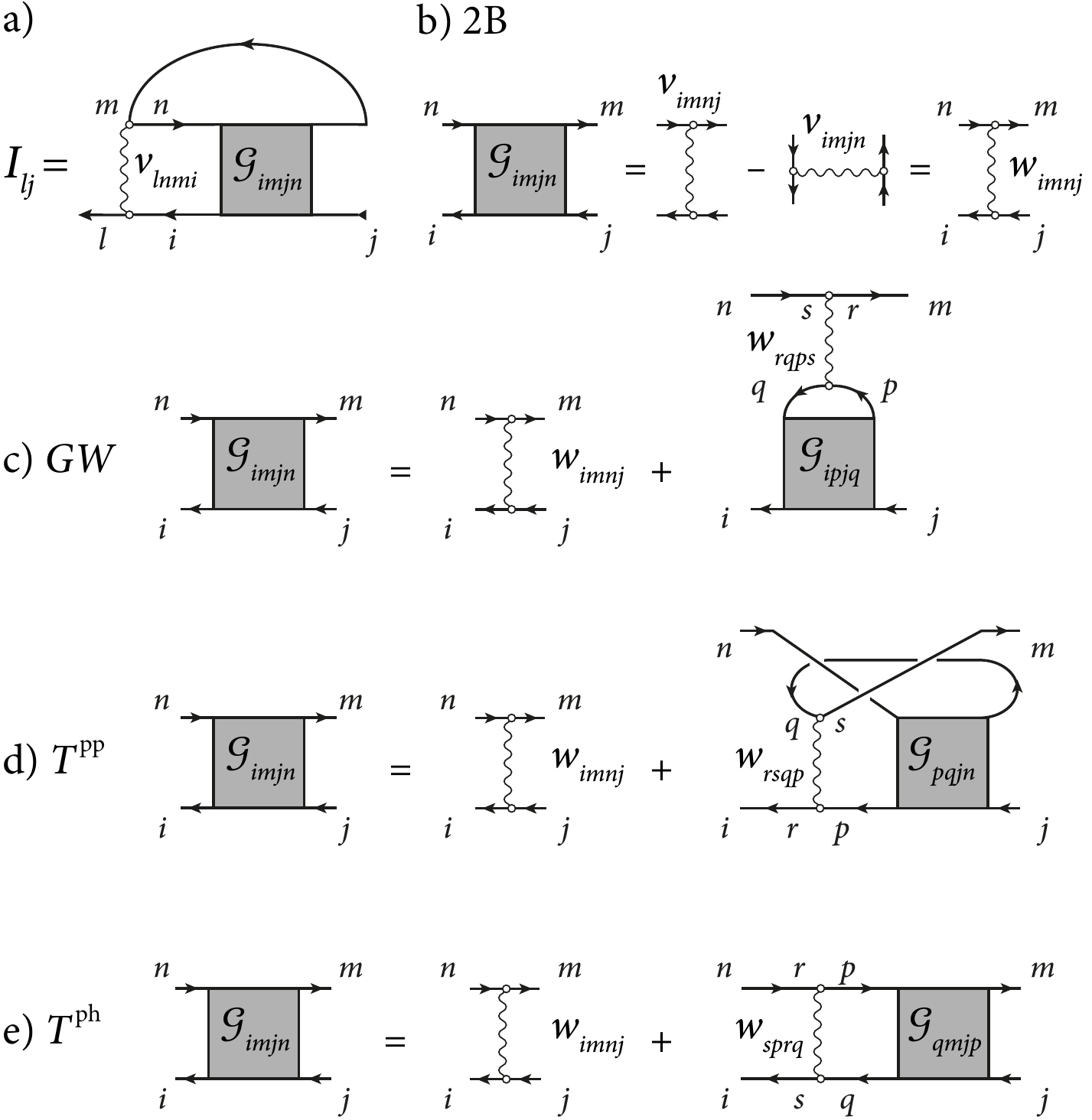}
\caption[]{Diagrammatic representation of the Eq.~\eqref{eq:Is} for the scattering
  term.\label{fig:Is}}
\end{figure}
In Eqs.~(\ref{eomrho}) and (\ref{hfgr})
\begin{align}
  h_{\text{HF},ij}(t)=h_{ij}(t)+\sum_{mn}[v_{imnj}-v_{imjn}]\rho^{<}_{nm}(t)
  \label{eq:hHF}
\end{align}
is the Hartree-Fock (HF) Hamiltonian, which is a functional of $\rho^{<}$. The so called
collision integral $I(t)$ in Eq.~(\ref{eomrho}) does therefore account for electronic
correlations and through the GKBA and Eq.~(\ref{hfgr}) it too is a functional of
$\rho^{<}$, see below. The ultimate goal for numerics is to compute the collision integral
in the most accurate and efficient fashion. Its exact form follows straightforwardly from
the first equation of the Martin-Schwinger hierarchy and it involves the two-particle
Green's function (2-GF) $\cG$ at equal times
\begin{align}
  I_{lj}(t)&=-\ii \sum_{imn} v_{lnmi}(t) \mathcal{G}_{imjn}(t).
  \label{eq:Is}
\end{align}
The diagrammatic expression of Eq.~(\ref{eq:Is}) is shown in Fig.~\ref{fig:Is}\,(a).

In this Section we evaluate the collision integral in the diagrammatic approximation $d=$
2B, see Fig.~\ref{fig:Is}\,(b), as well as $d=GW+(X)$, $T^{ph}+(X),\,T^{pp}+(X)$, see
Fig.~\ref{fig:Is}\,(c-e).  For the latter approximations the addition of exchange ($X$)
simply amounts in solving the Bethe-Salpeter equations of Fig.~\ref{fig:Is}\,(c-e) with an
interaction line $w_{imnj}=v_{imnj}-v_{imjn}$. In 2B the use of $w$ allows for writing the
direct and exchange diagrams in terms of a diagram only, see again
Fig.~\ref{fig:Is}\,(b). Depending on the approximation $d$ we find it convenient to
rewrite Eq.~(\ref{eq:Is}) in different, yet equivalent, forms
\begin{align}
  I_{lj}(t)&=-\ii \sum_{imn} v^{(d)}_{\substack{lm\\ in}}(t) 
  \mathcal{G}^{(d)}_{\substack{mj\\ ni}}(t)
  \label{eq:Is2}
\end{align}
where the relation between the {\em one-particle} 4-rank tensors $v$, $\mathcal{G}$ and
the {\em two-particle} 2-rank tensors $v^{(d)}$, $\mathcal{G}^{(d)}$ is provided in
Table~\ref{tab:1}.  To distinguish matrices (2-rank tensors) in the two-particle space
from matrices or tensors in the one-particle space we use bold letters for the former.  In
the following subsections we show that the GKBA expression for the 2-GF has the following
compact form for {\em all} approximations (omitting the dependence on $d$)
\begin{align}
\mcG(t)=\ii\int_{t_{0}}^{t}\!\!dt'\,\mPi^{R}(t,t')\mPsi(t')\mPi^{A}(t',t),
\label{eq:GKBA:G2}
\end{align}
where the initial time $t_{0}=0$, without any loss of generality. Thus $\mcG$ is the
integral of a product between ($d$-dependent) time-dependent matrices in the two-particle
space. The ($d$-dependent) matrix
\begin{align}
  \mPsi(t)\equiv \mrho^{>}(t)\mw(t)\mrho^{<}(t)-\mrho^{<}(t)\mw(t) \mrho^{>}(t)
  \label{Psidef}
\end{align}
is a simple product between the time-dependent matrices $\mrho^{\lessgtr}$ and $\mw$
defined in Table~\ref{tab:1}.  The ($d$-dependent) retarded propagator
$\mPi^{R}(t,t')=[\mPi^{A}(t',t)]^{\dag}$ satisfies for any $t>t'$ the differential
equation
\begin{align}
  \ii\frac{d}{dt}\mPi^{R}(t,t')=
  \big[\mh(t)+a \,\mw (t)\mrho^{\Delta}(t)\big]\mPi^{R}(t,t'),
\label{eq:Pi:EOM}
\end{align}
with the boundary condition 
\begin{align}
 \ii \mPi^{R}(t^{+},t)=\mathbb{1}\times
 \left\{\begin{array}{rl}
 -1, &\quad d={\rm 2B},\;GW;\\
  1, &\quad d=T^{ph},\;T^{pp}.
 \end{array}\right.
 \label{boundary}
\end{align}
The matrix $\mh$ as well as the constant $a$ are given in Table~\ref{tab:1} whereas
\begin{align}
  \mrho^\Delta(t)\equiv \mrho^{>}(t)-\mrho^{<}(t).
  \label{rhodelta}
\end{align}

The equation of motion for the 2-GF~\cite{schlunzen_achieving_2020,joost_g1-g2_2020}
follows directly from Eq.~(\ref{eq:Pi:EOM})
\begin{multline}
  \ii \frac{d}{dt}\mcG(t)=-\mPsi(t)+\left[\mh(t) 
   +a\mrho^\Delta(t)\mw(t)\right]\mcG(t)\\
  -\mcG(t) \left[\mh(t) +a\mw(t) 
    \mrho^\Delta(t)\right].
  \label{eq:G2:X:init}
\end{multline}
The coupled differential equations~\eqref{eomrho} and~\eqref{eq:G2:X:init} form the
essence of the NEGF+GKBA method for all the approximations in Tab.~\ref{tab:1}. The
numerical solution of these equations scales linearly with the propagation time.  The
concise derivation of such unifying formulation is made possible by the diagrammatic
structure of the 2-GF which takes into account only 2-particle correlations (in $p$-$p$ or
$p$-$h$ channels), see again Fig.~\ref{fig:Is}\,(c-e). We can then order the indices, see
Table~\ref{tab:1}, in such a way as to construct RPA-like equations in the respective
channels.  The contraction over the pair of indices translates in our notation to a matrix
product while the particle permutation symmetry is taken into account by the constant
$a$. The 2B approximation is the lowest order term of all the correlated methods, $GW$ and
$T$-matrix, when exchange is added. Accordingly, the 2B equations (now expressed in the
$GW$ index convention) can be equivalently formulated using the $T$-matrix index
conventions -- this point is expanded in Section~\ref{2Bder}. We also observe that
\begin{align}
  \mcG^\dagger&=\mcG,&\mv^\dagger&=\mv,&\mw^\dagger&=\mw,
  &\mh^\dagger&=\mh,&{\mrho^{\gtrless}}^{\dagger}&=\mrho^{\gtrless},
  \label{eq:sym:2}
\end{align}
as it follows directly from the symmetry properties 
\begin{subequations}
\begin{align}
  v_{1234}&=v_{4321}^*=v_{2143},\\
  \cG_{1234}&=\cG_{3421}^*=\cG_{2143}.
\end{align}
\end{subequations}

In the remainder of the Section we present the derivation of Eq.~(\ref{eq:GKBA:G2}). We
point out, however, that it is not necessary to go through the derivation in order to
follow the Faddeev approximation in Section~\ref{sec:2h-1p}.

\subsection{Second Born approximation}
\label{2Bder}
\begin{figure}[t!]
\centering  \includegraphics[width=0.9\columnwidth]{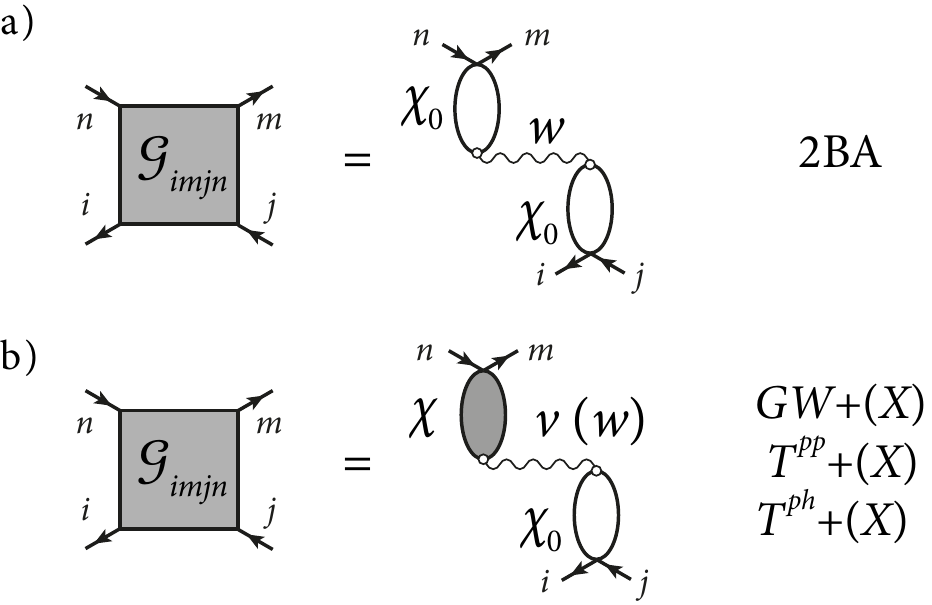}
\caption[]{Diagrammatic representation of Eq.~\eqref{eq:G2:0:2ba} (a) and
  Eq.~\eqref{eq:G2:0} (b) for the two-particle Green's function in terms of the RPA $\mchi$
  and noninteracting $\mchi^{0}$ response functions. \label{fig:G2:chi}}
\end{figure}

Let us start with the simplest case, where the collision integral is given by its
second-order (in $v$) expression\,---\,therefore the name.  The equal-time 2-GF can be
expressed as the convolution of two response functions $\mchi^{0}$, see
Fig.~\ref{fig:G2:chi}\,(a),
\be
\mcG(t)=-\ii\int_{0}^{t}\!\!dt'\Big\{
 \mchi^{0,>}(t,t')\mw(t')\mchi^{0,<}(t',t) -(>\leftrightarrow<)
\Big\}. \label{eq:G2:0:2ba}
\ee
As already pointed out, in 2B  there is a freedom in selecting the 
index convention. Let us define three different response functions 
as matrices in the two-particle space
\begin{align}
  \chi^{0,\lessgtr}_{\begin{subarray}{c}13\\24\end{subarray}}(t,t')=
  \ii\left\{\begin{array} {ll}
  -G^{\lessgtr}_{13}(t,t')G^{\gtrless}_{42}(t',t),& \quad GW\\[6pt]
  +G^{\lessgtr}_{13}(t,t')G^{\lessgtr}_{24}(t,t'),& \quad T^{pp}\\[6pt]
  +G_{13}^{\lessgtr}(t,t')G^{\gtrless}_{42}(t',t),& \quad T^{ph}
  \end{array}
  \right..
\label{eq:def:chi0}
\end{align}
Then the collision integral in Eq.~(\ref{eq:Is2}) does not change if we consistently use
for $\mv$, $\mchi^{0,\lessgtr}$ and $\mw$ the same index convention as described in
Table~\ref{tab:1}.

Evaluating the noninteracting response functions with the GKBA in Eq.~(\ref{eq:gkba}) we
find
\be
\mchi^{0,\lessgtr}(t,t')=\mP^R(t,t')\mrho^{\lessgtr}(t')-\mrho^{\lessgtr}(t)\mP^A(t,t'),
\label{eq:gkba:chi0}
\ee
where, depending on the approximation $d=GW,\, T^{pp},\, T^{ph}$,
\begin{align}
  \mP^{R}_{\begin{subarray}{c}13\\24\end{subarray}}(t,t')&=
  \begin{cases}
    +\ii G_{13}^R(t,t') G_{42}^A(t',t),&GW\\[4pt]
    +\ii G_{13}^R(t,t') G_{24}^R(t',t),&T^{pp}\\[4pt]
    -\ii G_{13}^R(t,t') G_{42}^A(t',t),&T^{ph}
  \end{cases},\label{eq:PR}
\end{align}
\begin{align}
  \mP^{A}(t,t')&=\left[\mP^{R}(t',t)\right]^\dagger,
  \label{avsr}
\end{align}
and the matrices $\mrho^{>}$ and $\mrho^{<}$ are defined in Table~\ref{tab:1} for each
diagrammatic approximation.  Taking into account that $t'\le t$ in
Eq.~\eqref{eq:G2:0:2ba}, substituting Eqs.~\eqref{eq:gkba:chi0} into it we arrive at
\be
\mcG(t)=\ii\int_{0}^{t}\!\!dt'\,
\mP^R(t,t')\mPsi(t')\mP^{A}(t',t).\label{eq:GKBA:G2:2ba}
\ee
Comparing this result with Eq.~(\ref{eq:GKBA:G2}) we are left to prove that $\mP^R$
satisfies Eq.~(\ref{eq:Pi:EOM}) with boundary condition in Eq.~(\ref{boundary}). The
equation of motion for $\mP^R$ follows from the equation of motion of the
retarded/advanced Green's functions. According to Eq.~\eqref{hfgr} we have (repeated
indices are summed over)
\begin{subequations}
\label{eq:eom:G:RA}
\begin{align}
  \ii\frac{d}{dt}G^{R/A}_{ma}(t,t')&=h_{mc}^\text{HF}(t)G^{R/A}_{ca}(t,t')+
  \delta_{ma}\delta(t-t'),\\
  -\ii\frac{d}{dt'}G^{R/A}_{ma}(t,t')&=G^{R/A}_{mc}(t,t')h_{ca}^\text{HF}(t')+
  \delta_{ma}\delta(t-t').
\end{align}
\end{subequations}
By defining the matrix $\mh$ in the two-particle space according to 
Table~\ref{tab:1} we can then write for all cases and for $t>t'$
\begin{align}
  \ii\frac{d}{dt}\mP^R(t,t')&=\mh(t)\mP^R(t,t'),
\label{eq:EOM:P}
\end{align}
which coincides with Eq.~(\ref{eq:Pi:EOM}) when $d=$ 2B since in this case $a=0$.  The
initial conditions for $\mP^{R}$ can likewise be obtained by taking the equal time-limit
of Eq.~\eqref{eq:PR} and by using $G^R(t^+,t)=-\ii$ and $G^A(t,t^+)=\ii$:
\begin{align}
  \ii \mP^{R}(t^+,t)& =\mathbb{1}\times
  \begin{cases}
    -1 & GW;\\
    +1 & T^{pp},\,T^{ph}.
  \end{cases}
  \label{eq:PR:0}
\end{align}
Using the $GW$ index convention (this is the convention used in 
Tab.~\ref{tab:1} for 2B) we find the boundary condition of 
Eq.~(\ref{boundary}).

\subsection{$GW$ and $T$-matrices approximation}

Higher-order diagrammatic approximations for $\mcG$ such as $GW$ and $T$-matrices with
exchange require the notion of the \emph{RPA response} functions $\mchi$ as depicted in
Fig.~\ref{fig:G2:chi}\,(b). For all the cases we can write
\be
\mcG(t)=-\ii\int^{t}_{0}\!\!dt'\Big\{
 \mchi^{>}(t,t')\mw(t')\mchi^{0,<}(t',t)\\
 -(>\leftrightarrow<)\Big\},\label{eq:G2:0}
\ee
where the non-interacting $\mchi^{0}$ has been defined in Eq.~(\ref{eq:def:chi0}).  To
recover the more standard $GW$ and $T$-matrix approximations we can simply replace $\mw$
with $\mv$ in Eq.~(\ref{eq:G2:0}) and in the RPA equation for $\mchi$. We can also
consider the exchange-only version of these approximations; in this case the replacement
is $\mw\to\mw-\mv$.  To reduce the voluminousness of the equations we introduce the
two-time function $\mw(t,t')=\mw(t)\delta(t-t')$ and the shorthand notation
\begin{align}
    [a\cdot b](t,t')=\int\! d\bar{t}\,a(t,\bar{t})b(\bar{t},t').
\end{align}
The Langreth rules then imply, see Appendix~\ref{sec:app:id},
\begin{align}
\mchi^{\lessgtr}=
(\md+\mchi^R\cdot \mw)\cdot\mchi^{0,\lessgtr}\cdot(\mw\cdot\mchi^A+\md)
\label{chilessgtr}
\end{align}
with
\begin{subequations}
  \label{eq:rpa}
\begin{align}
  \mchi^{R/A}&=\mchi^{0,R/A}+\mchi^{0,R/A}\cdot \mw\cdot\mchi^{R/A},\label{eq:rpa:1}\\
  &=\mchi^{0,R/A}+\mchi^{R/A}\cdot\mw\cdot\mchi^{0,R/A}.\label{eq:rpa:2}
\end{align}
\end{subequations}
In the $GW$ case $\mchi$ is well-known as the density-density response
function, of high relevance for the optical properties.  
Let us work out the expression of $\mchi$ when the Green's function is 
evaluated using the GKBA. 

By definition $\mchi^{0,R/A}(t,t')=\pm \theta(\pm t\mp 
t')[\mchi^{0,>}(t,t')-\mchi^{0,<}(t,t')]$. Hence from Eq.~(\ref{eq:gkba:chi0})
\begin{subequations}
\begin{align}
  \mchi^{0,R}(t,t')&=\mP^R(t,t')\mrho^\Delta(t'),\\
  \mchi^{0,A}(t,t')&=\mrho^\Delta(t)\mP^A(t,t'),
\end{align}
\label{eq:def:P}%
\end{subequations}
where $\mrho^\Delta$ has been defined in Eq.~(\ref{rhodelta}).  Inserting this result into
Eqs.~(\ref{eq:rpa}) we find
\begin{subequations}
\begin{align}
  \mchi^{R}(t,t')&=\mPi^R(t,t')\mrho^\Delta(t'),\\
  \mchi^{A}(t,t')&=\mrho^\Delta(t)\mPi^A(t,t'),
\end{align}
\label{fullchiRA}%
\end{subequations}
where $\mPi^{R/A}$ satisfy the RPA equations
\begin{subequations}
\begin{align}
  \mPi^{R/A}&=\mP^{R/A}+ \mPi^{R/A}\cdot\mrho^\Delta\mw\cdot\mP^{R/A}\\
  & =\mP^{R/A}+ \mP^{R/A}\cdot\mrho^\Delta\mw\cdot\mPi^{R/A}.
\end{align}
\label{eq:rpa:PiR}%
\end{subequations}
In Eqs.~(\ref{eq:rpa:PiR}) the quantities $[\mw\mrho^\Delta](t,t')\equiv
\mw(t,t')\mrho^\Delta(t')$ and $[\mrho^\Delta\mw](t,t')\equiv \mrho^\Delta(t)\mw(t,t')$.
Notice that $\mP$ and $\mPi$, unlike the response functions $\mchi^{0}$ and $\mchi$, are
auxiliary quantities that cannot be written as operator averages, i.\,e., they are not
correlators.

Using the GKBA for $\mchi^{0,\lessgtr}$ [Eq.~\eqref{eq:gkba:chi0}] and the GKBA for
$\mchi^{R/A}$ [Eq.~\eqref{fullchiRA}] in Eq.~(\ref{chilessgtr}) we obtain a rather concise
form for $\mchi^{\lessgtr}$ that can be paralleled with Eq.~\eqref{eq:gkba:chi0}
\begin{align}
  \mchi^{\lessgtr}
  &=\big(\md+\mPi^R\cdot\mrho^\Delta\mw\big)\cdot
  \big(\mP^R\mrho^{\lessgtr}-\mrho^{\lessgtr}\mP^A\big)
  \cdot\big(\mw\mrho^\Delta\cdot\mPi^A+\md\big)\nn\\
  &=\mPi^R\mrho^{\lessgtr}\cdot\big(\md+\mw\mrho^\Delta\cdot\mPi^A\big)-
  \big(\mPi^R\cdot\mrho^\Delta\mw+\md\big)\cdot\mrho^{\lessgtr}\mPi^A.
  \label{eq:chi:gtrless}
\end{align}
Inserting now Eq.~(\ref{eq:gkba:chi0}) for $\mchi^{0,\lessgtr}$ and
Eq.~(\ref{eq:chi:gtrless}) for $\mchi^{\lessgtr}$ into Eq.~(\ref{eq:G2:0}), and using
again the RPA equations~\eqref{eq:rpa:PiR} that relate $\mPi$ and $\mP$, the general
result for $\mcG(t)$ in Eq.~(\ref{eq:GKBA:G2}) follows. We are then left to prove that
$\mPi^{R}$ satisfies Eq.~(\ref{eq:Pi:EOM}) with boundary condition (\ref{boundary}).

This can be achieved by first observing that the property (\ref{avsr}) transfers directly
to $\mPi^R$ and $\mPi^A$ through the RPA equations. The retarded/advanced nature of the
$\mPi$ functions further implies that $\ii \mPi^R(t^{+},t)=\ii \mP^R(t^{+},t)$\,---\,hence
the boundary condition (\ref{boundary}). Finally, the equation of motion (\ref{eq:Pi:EOM})
follows by differentiating Eq.~\eqref{eq:rpa:PiR} and using the equation of motion
(\ref{eq:EOM:P}) for $\mP^R(t,t')$ with the boundary condition Eq.~(\ref{eq:PR:0}).
\section{The  Faddeev approximation in GKBA\label{sec:2h-1p}}

The main objective of this work is to develop an accurate and efficient approximation
scheme to simulate the electron dynamics of organic molecules induced by a weak XUV pulse.
In these systems {\em ground-state} electronic correlations are rather weak and the
unperturbed many-body state is well approximated by a Slater determinant of HF
wavefunctions. The weak XUV pulse extracts one electron from the inner valence states
causing hole migration.  In a simplifying picture the hole can either move ``freely'' in
the space of the originally occupied ($\mathcal{O}$) HF molecular orbitals (MOs) or
scatter with an electron in one of the unoccupied ($\mathcal{V}$) HF MOs thereby creating
another particle-hole pair. The ``free'' motion is captured by a time-dependent HF
treatment (which amounts to discard the collision integral). The second process,
henceforth called the {\em shake-up} process, is instead triggered by the Coulomb
integrals $v_{imnj}$ with only one index in the unoccupied sector. As we shall see these
processes require a nonperturbative treatment of three-particle correlators for the
evaluation of the collision integral. In the reminder of this work the spin indices are
explicitly spelled out for clarity.

\subsection{Shake-up effects}

We recall that the Hamiltonian is invariant under spin-flip and that the initial state is
spin-compensated; it is therefore sufficient to calculate the up-up component of the
density matrix since $\rho_{i\s_{1}j\s_{2}}=\d_{\s_{1}\s_{2}}\rho_{ij}$. Let us denote by
$v^{s}$ the shake-up Coulomb tensor defined as $v^{s}=v$ if only one index belongs to the
unoccupied sector and three other indices are distinct occupied ones, and $v^{s}=0$
otherwise. We can then write the shake-up contribution to the collision integral as
\be
I_{lj}^{s}(t)\equiv I_{l\ua j\ua}^{s}(t)=-\ii
\sum_{\substack{imn\\\s}} v_{lnmi}^{s} \cG_{i\ua m\s j\ua n\s}(t),
\label{eq:I:s:0}
\ee
where $l$, $j$, $i$, $m$, $n$ are the orbital indices. In deriving this equation we have
made use of the spin-structure of the Coulomb tensor, see Eq.~(\ref{vspin}).  After the
XUV pulse has passed through the molecule, the equal-time 2-GF is given by
\be \cG_{i\ua m\s
  j\ua n\s}(t)=\frac{1}{\ii^{2}}\bra\varphi(t)| \hat{d}_{n\s}^\dagger \hat{d}_{j\ua}^\dagger
\hat{d}_{i\ua} \hat{d}_{m\s} |\varphi(t)\ket,\
\label{callGxsu}
\ee
where $|\varphi(t)\ket=e^{-\ii \hat H t}|\varphi\ket$ and $|\varphi\ket$ is the state of
the molecule just after the pulse.  This state differs from the HF ground state
$|\phi_{\rm HF}\ket$ since it contains a small component in the cationic space:
$|\varphi\ket=(1+\sum_{k\s}\alpha_{k}\hat{d}_{k\s})|\phi_{\rm HF}\ket$ where the
coefficients $\alpha_{k}\ll 1$ are linear in the electric field of the XUV pulse.  We
intend to approximate $\cG$ to the lowest order in the shake-up transition amplitudes
$v^{s}$ while still treating nonperturbatively $2h$-$1p$ correlation effects.  For this
purpose we write the full Coulomb tensor as $v=v^{s}+v'$ and retain in $v'$ only the
two-index direct and exchange integrals. This means that
\be v'_{imnj}=
\d_{ij}\d_{mn}v_{immi}+\d_{in}\d_{mj}v_{imim}.
\ee
This selection of Coulomb integrals is dictated by the fact that only direct and exchange
terms contribute to the energy of $2h$-$1p$ states, see below. The full Hamiltonian
appearing in Eq.~(\ref{callGxsu}) is then approximated as
\be \hat{H}\simeq \hat{H}'+\hat{H}^{s}_{\rm int},
\label{approxham}
\ee
where, see Eq.~(\ref{ham}),
\be
\hat H'=\sum_{\substack{ij\\\s}} h_{ij}\hat{d}_{i\s}^\dagger 
\hat{d}_{j\s}
+\frac12\sum_{\substack{ijmn\\\s\s'}}v'_{ijmn}
  \hat{d}_{i\s}^\dagger \hat{d}_{j\s'}^\dagger 
  \hat{d}_{m\s'}\hat{d}_{n\s},
\ee
and
\be
\hat{H}^{s}_{\rm int}=\frac12\sum_{\substack{ijmn\\\s\s'}}v^{s}_{ijmn}
  \hat{d}_{i\s}^\dagger \hat{d}_{j\s'}^\dagger 
  \hat{d}_{m\s'}\hat{d}_{n\s}.
\ee
Notice that no double counting occurs in Eq.~(\ref{approxham}) since $v^{s}$ has only one
index in the $\mathcal{V}$-sector and $v'$ has orbital indices equal in pairs. We also
remind the reader that $h_\text{HF}$ is always evaluated with the full Coulomb tensor
[cf. Eq~\eqref{eq:hHF}].

Approximating $\hat{H}$ like in Eq.~(\ref{approxham}) and expanding Eq.~(\ref{callGxsu})
to first order in $v^{s}$ we obtain
\be 
\cG_{i\ua m\s j\ua n\s}(t)=A^{\s}_{imjn}(t)+A_{njmi}^{\s\ast}(t),
\label{g2decomp}
\ee
with
\begin{align}
A^{\s}_{imjn}(t)
  =\frac{1}{2\ii^3}\int_{0}^{t}d\bt\sum_{\substack{pqrs\\\s_{1}\s_{2}}} v^{s}_{pqrs}
  {G_{4}}_{m\s i\ua j\ua n\s}^{p\s_{1}q\s_{2}r\s_{2}s\s_{1}}(t,\bt),
\end{align}  
and
\begin{multline}
{G_{4}}_{m\s\, i\s'\, j\s'\, n\s}^{p\s_{1}q\s_{2}r\s_{2}s\s_{1}}(t,\bt)
\equiv
  \langle\varphi(t)|\hat{d}_{n\s}^\dagger \hat{d}_{j\ua}^\dagger 
\hat{d}_{i\ua} \hat{d}_{m\s} 
  e^{-\ii \hat{H}' (t-\bt)}
  \\
  \hat{d}_{p\s_{1}}^\dagger \hat{d}_{q\s_{2}}^\dagger \hat{d}_{r\s_{2}} 
  \hat{d}_{s\s_{1}}|\varphi(\bt)\rangle.
\end{multline}
Since shake-up processes have been removed from $\hat{H}'$ and $|\varphi\ket$ has no
electrons in the unoccupied sector, we conclude that the indices $r$ and $s\in
\mathcal{O}$ (belong to the occupied sector). This also implies that either $p$ or $q\in
\mathcal{V}$ for otherwise $v^{s}_{pqrs}$ would vanish. Therefore $A_{imjn}$ is
nonvanishing only if the indices $m$ or $i\in \mathcal{V}$ (are unoccupied).  Shake-up
scatterings are then of two kinds: (i) initial hole in $i$ and final $2h$-$1p$ in the
states $nj$-$m$ or (ii) initial hole in $m$ and final $2h$-$1p$ in the states $nj$-$i$.
To fully account for the three-particle correlations we make the following approximation
\begin{align}
{G_{4}}_{m\s\, i\ua\, j\ua\, n\s}^{p\s_{1}q\s_{2}r\s_{2}s\s_{1}}(t,\bt)
\simeq -\bar{f}_m\Big\{\d_{\s_{2}\ua}G_{iq}^>(t,\bt)
{G_{3}^{<}}_{m\s j\ua n\s}^{p\s_{1}r\ua s\s_{1}}(t,\bt)
  \nonumber \\
  -\d_{\s_{1}\ua}G_{ip}^>(t,\bt)
{G_{3}^{<}}_{m\s j\ua n\s}^{q\s_{2}r\ua s\s_{1}}(t,\bt)\Big\}
\nonumber\\
  -\bar{f}_i\Big\{\d_{\s\s_{1}}G_{mp}^>(t,\bt){G_{3}^{<}}^{q\s_{2}r\s_{2}s\s}_{i\ua j\ua n\s}(t,\bt)
  \nn\\
  -\d_{\s\s_{2}}G_{mq}^>(t,\bt){G_{3}^{<}}^{p\s_{1}r\s s\s_{1}}_{i\ua j\ua n\s}(t,\bt)\Big\},
  \label{eq:OO}
\end{align}  
where $\bar{f}_{m}=1$ if $m$ is initially occupied and zero otherwise. In
Eq.~(\ref{eq:OO}) we have introduced the central object of the Faddeev approximation,
i.\,e., the $2h$-$1p$ GF
\begin{multline}
\ii {G_{3}^{<}}_{m\s_{1}j\s_{2}n\s_{3}}^{p\s'_{1}r\s'_{2}s\s'_{3}}(t,\bt)\equiv
  \langle\varphi(t)|\hat{d}_{n\s_{3}}^\dagger \hat{d}_{j\s_{2}}^\dagger \hat{d}_{m\s_{1}}
   e^{-\ii\hat{H}'(t-\bt)}
   \\
   \hat{d}_{p\s'_{1}}^\dagger \hat{d}_{r\s'_{2}} 
   \hat{d}_{s\s'_{3}}|\varphi(\bt)\rangle.
   \label{eq:G2h1p:def}
\end{multline}

\subsection{GKBA for the $2h$-$1p$ Green's function}

From Eq.~(\ref{eq:OO}) we see that the GKBA for the lesser and greater Green's function is
not sufficient for closing the equation of motion (\ref{eomrho}) since $G_{3}$ is not an
explicit functional of the density matrix.  We pursue here the idea of extending the GKBA
to higher order Green's functions and propose the following form for ${G_{3}^{<}}(t,\bt)$
when $\t=t-\bt>0$
\begin{multline}
{G_{3}^{<}}_{m\s_{1}j\s_{2}n\s_{3}}^{p\s'_{1}r\s'_{2}s\s'_{3}}(t,\bt)
=\left[{G_{3}^{R}}_{\s_{1}\s_{2}\s_{3}}^{\s'_{1}\s'_{2}\s'_{3}}(t,\bt)\right]_{mjn}
\times \rho^{>}_{mp}(\bt)\rho^{<}_{rj}(\bt)\rho^{<}_{sn}(\bt)
\\
-\left[{G_{3}^{R}}_{\s_{1}\s_{2}\s_{3}}^{\s'_{1}\s'_{3}\s'_{2}}(t,\bt)\right]_{mjn}
\times \rho^{>}_{mp}(\bt)\rho^{<}_{rn}(\bt)\rho^{<}_{sj}(\bt).
\label{eq:G2h1p:fct}
\end{multline}
The motivation for this Ansatz is that the evolution operator $e^{-\ii\hat{H}'(t-\bt)}$
evolves the bra state from time $t$ to $\bt$, whereby the scattering takes place on the
same subset of $1p$-$2h$ states, possibly changing the spin. The whole argument is
detailed in App.~\ref{gkbaintroapp}.  Evaluating $G_{4}$ in Eq.~(\ref{eq:OO}) using the
GKBA expressions for $G^{>}$ and $G^{<}_{3}$ we obtain
\begin{align}
A^{\s}_{imjn}(t)=-\int_{0}^{t}\!d\bt\;\Psi_{imnj}(\bt) \sum_{\s'}
\Big\{&\bar{f}_{m}e^{-\ii\e_{i}\t}\left[{G_{3}^{R}}_{\s\,\ua\,\s}^{\s'\ua\s'}(t,\bt)\right]_{mjn}
\nn\\
-&\bar{f}_{m}e^{-\ii\e_{i}\t}\left[{G_{3}^{R}}_{\s\,\ua\,\s}^{\s'\s'\ua}(t,\bt)\right]_{mjn}
\nn\\
+&\bar{f}_{i}e^{-\ii\e_{m}\t}\left[{G_{3}^{R}}_{\ua\,\ua\,\s}^{\s'\s'\s}(t,\bt)\right]_{ijn}
\nn\\
-&\bar{f}_{i}e^{-\ii\e_{m}\t}\left[{G_{3}^{R}}_{\ua\,\ua\,\s}^{\s'\s\s'}(t,\bt)\right]_{ijn}\Big\},
\label{Agkba}
\end{align}
where we have defined
\be
\Psi_{imnj}(t)\equiv 
\sum_{pqrs}v^{s}_{pqrs}\rho^{>}_{mp}(t)\rho^{>}_{iq}(t)\rho^{<}_{rj}(t)\rho^{<}_{sn}(t).
\ee
In Eq.~(\ref{Agkba}) we have also used that the XUV pulse is weak (only single-photon
ionization events are considered) and hence the retarded Green's function in
Eq.~(\ref{hfgr}) is well approximated by the equilibrium expression
\begin{align}
G^{R}_{ip}(t,\bt)=-\ii\d_{ip}\theta(\t)e^{-\ii\e_{i}\t},
\label{hfgreq}
\end{align}
where $\e_{i}$ is the eigenvalue of the equilibrium single-particle HF Hamiltonian.

\subsection{Faddev approximation for the $1p$-$2h$ propagator}
\label{sec:spin}
\begin{figure}[t!]
\centering\includegraphics[width=\columnwidth]{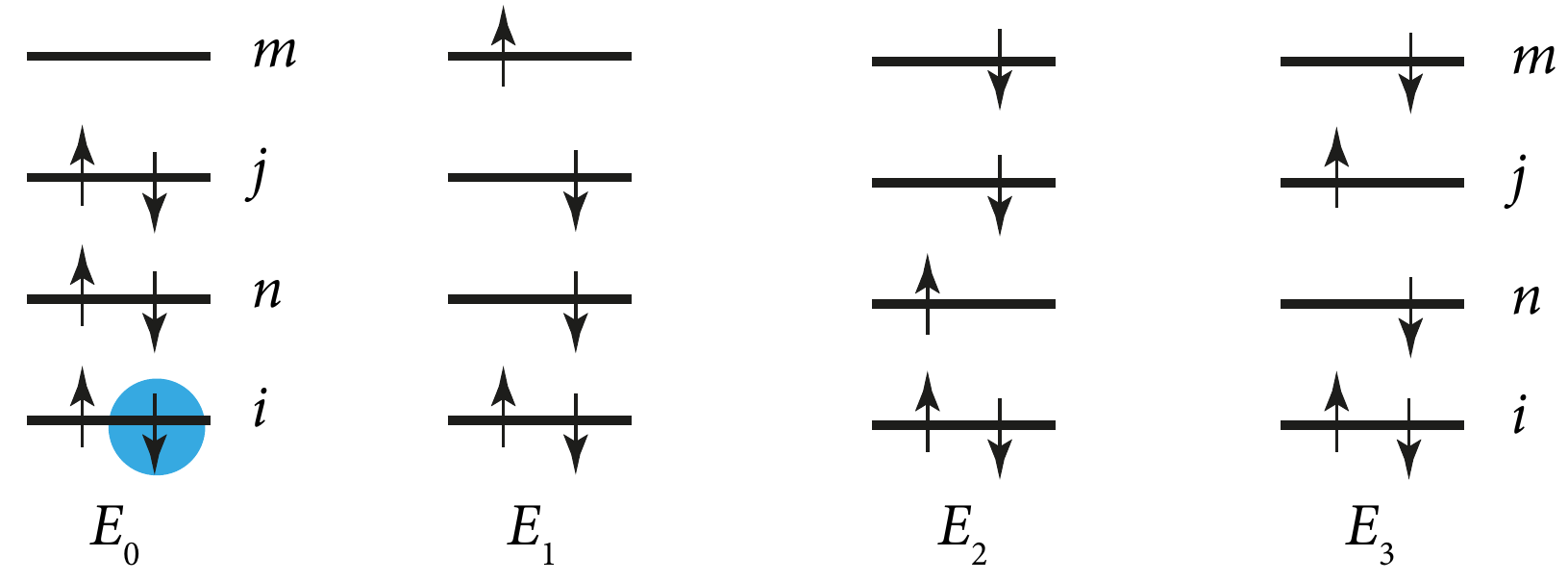}
\caption[]{\small Ground state configuration of the neutral system and the cationic Slater
  determinants of interest. A hole created in the state $i$ (denoted here as a blue
  circle) may decay by virtue of the $1h\rightarrow 2h$-$1p$ scattering into one of the six
  states defined by Eqs.~\eqref{eq:full:subspace} (only three are shown).\label{fig:0}}
\end{figure}
Now we come to the most interesting non-perturbative aspect concerning the evaluation of
$G_3^{R}$. This object is the evolution operator on a fixed subspace of three orbitals as
only spin can change, see Fig.~\ref{fig:0}.  To the best of our knowledge none of the
established diagrammatic approximations can deal with this scenario in nonequilibrium
situations.  Accounting for $2h$-$1p$ correlations is, however, mandatory for a good
description of the shake-up processes.

The states involved in the first two terms of Eq.~(\ref{Agkba}) can be grouped into two
triplets differing only by a spin flip
\begin{subequations}
  \label{eq:full:subspace}
\begin{align}
  |1_{\ua}\rangle&\equiv\hat{d}^{\dag}_{m\ua}\hat{d}_{j\ua}\hat{d}_{n\ua}|\phi_{\rm HF}\ket, &
  |1_{\da}\rangle&\equiv\hat{d}^{\dag}_{m\da}\hat{d}_{j\da}\hat{d}_{n\da}|\phi_{\rm HF}\ket;\\
  |2_{\ua}\rangle&\equiv\hat{d}^{\dag}_{m\da}\hat{d}_{j\da}\hat{d}_{n\ua}|\phi_{\rm HF}\ket,&
  |2_{\da}\rangle&\equiv\hat{d}^{\dag}_{m\ua}\hat{d}_{j\ua}\hat{d}_{n\da}|\phi_{\rm HF}\ket;\\
  |3_{\ua}\rangle&\equiv\hat{d}^{\dag}_{m\da}\hat{d}_{j\ua}\hat{d}_{n\da}|\phi_{\rm HF}\ket,&  
  |3_{\da}\rangle&\equiv\hat{d}^{\dag}_{m\ua}\hat{d}_{j\da}\hat{d}_{n\ua}|\phi_{\rm HF}\ket.
\end{align}
\end{subequations}
For any $\a=1,2,3$ the states $|\a_{\ua}\rangle$ have spin-projection $S_z=-1/2$,
whereas the states $|\a_{\da}\rangle$ have $S_z=1/2$. Therefore, the matrix 
representing the Hamiltonian $\hat{H}'$ in the subspace 
$\{|\a_{\s}\ket,\a=1,2,3;\s=\ua,\da\}$  has a 
block-diagonal form with two identical $3\times 3$ blocks. Denoting by 
$\mathfrak{h}_{mjn;\a\b}\equiv \bra \a_{\ua}|\hat{H}'| \b_{\ua}\ket$ the matrix 
elements of the $3\times 3$ block one finds
\be
\mathfrak{h}_{mjn}=\left(\begin{array}{ccc}
E_{mjn}^1 & v_{mj}^{x} & v_{mn}^x\\
v_{mj}^{x}& E_{mjn}^2 & -v_{jn}^x\\
v_{mn}^x  & -v_{jn}^x &E_{mjn}^3
\end{array}
\right),
\label{eq:h:eff}
\ee
with
\begin{subequations}
\begin{align}
  E_{mjn}^1&=\epsilon_m-\epsilon_j-\epsilon_n-w_{mj}-w_{mn}+w_{jn},\\
  E_{mjn}^2&=\epsilon_m-\epsilon_j-\epsilon_n-w_{mj}-v_{mn}^d+v_{jn}^d,\\
  E_{mjn}^3&=\epsilon_m-\epsilon_j-\epsilon_n-v_{mj}^d-w_{mn}+v_{jn}^d,
\end{align}
\end{subequations}
and the direct, exchange and antisymmetrized Coulomb matrix elements read
\begin{align}
  v^d_{\mu\nu}&=v_{\mu\nu\nu\mu},&
  v^x_{\mu\nu}&=v_{\mu\nu\mu\nu},&
  w_{\mu\nu}&=v^d_{\mu\nu}-v^x_{\mu\nu}.
  \label{eq:v:2index}
\end{align}
We notice that using the full Hamiltonian $H$ in place of $H'$ in the definition of
$\mathfrak{h}$ would not change the result; this justifies the splitting of Coulomb
integrals in Eq.~\eqref{approxham}.  Our approximation for the retarded $2h$-$1p$
propagators appearing in Eq.~(\ref{Agkba}) is then (omitting the subscript $mjn$)
\be
\left(\begin{array}{ccc}
  {G_{3}^{R}}_{\ua\ua\ua}^{\ua\ua\ua}(t,\bt)&{G_{3}^{R}}_{\ua\ua\ua}^{\da\da\ua}(t,\bt)&{G_{3}^{R}}_{\da\ua\da}^{\ua\ua\ua}(t,\bt)
  \\[6pt]
    {G_{3}^{R}}_{\ua\ua\da}^{\da\da\da}(t,\bt)&{G_{3}^{R}}_{\ua\ua\da}^{\ua\ua\da}(t,\bt)&{G_{3}^{R}}_{\ua\ua\da}^{\ua\da\ua}(t,\bt)
    \\[6pt]
      {G_{3}^{R}}_{\da\ua\da}^{\ua\ua\ua}(t,\bt)&{G_{3}^{R}}_{\da\ua\da}^{\da\da\ua}(t,\bt)&{G_{3}^{R}}_{\da\ua\da}^{\da\ua\da}(t,\bt)
\end{array}
\right)=-\ii\theta(\t)e^{-\ii\mathfrak{h}(\t)}.
\label{faddeevGr}
\ee

It is important to comment on the relation between the Faddeev approximation and the
conventional approaches discussed in Section~\ref{sec:derivation}. In the $GW$
approximation as well as in the $T$-matrix approximation in the $ph$ channel one of the
holes is a mere spectator and only the scattering between the particle and the other hole
is treated to infinite order. Similarly, in the $T$-matrix approximation in the $pp$
channel the particle is a mere spectator while the scattering between the two holes is
treated nonperturbatively. It is possible to recover these approximations by retaining in
the $\mathfrak{h}_{mjn}$ matrix only the direct and exchange Coulomb integrals of the
considered channel. In particular, the $GW$ approximation is recovered by retaining only
$v^{d}_{mn}$ and $v^{x}_{mn}$, the $T$-matrix approximation in the $ph$ channel is
recovered by retaining only $v^{d}_{mj}$ and $v^{x}_{mj}$ and the $T$-matrix approximation
in the $pp$ channel is recovered by retaining only $v^{d}_{nj}$ and $v^{x}_{nj}$.

\subsection{Working formulas}

According to Eq.~(\ref{g2decomp}) the collision integral can be written as
\begin{align}
  I^{s}_{l j}(t)&=-\ii \sum_{imn}v^{s}_{lnmi}
  \left(A_{imjn}+A^{\ast}_{jnim}\right),
  \label{eq:I:s}
\end{align}
where $A_{imjn}(t)\equiv\sum_{\s}A^{\s}_{imjn}(t)$.  Let $\Omega^{\l}$ and $Y^{\l}$ be the
eigenvalues and the eigenvectors of the $3\times3$ Hamiltonian in Eq.~\eqref{eq:h:eff}:
$\mathfrak{h}_{mjn}Y^{\l}_{mjn}=\Omega_{mjn}^{\l}Y_{mjn}^{\l}$.  Using the spectral
decomposition
\be
[e^{-\ii\mathfrak{h}\t}]_{\a\b}=\sum_{\l}e^{-\ii\Omega^{\l}\t}Y^{\l\a}Y^{\l\b}
\ee
to write the $2h$-$1p$ propagator we obtain the following expression for $A_{imjn}$
\begin{align}
  A_{imjn}(t)&=\textstyle\sum_\l \left(P_{imjn}^{\l}(t)+Q_{imjn}^{\l}(t)\right),
  \label{eq:A:1p2h}
\end{align}
where $P_{imjn}^{\l}(t)$ and $Q_{imjn}^{\l}(t)$ are obtained from the solution of ODEs:
\begin{subequations}
  \label{eq:PQ}
\begin{align}
  \ii\frac{d}{dt}P_{imjn}^{\l}(t)&=\bar{f}_m
  \Big[\big(Y_{mjn}^{\l1}+Y_{mjn}^{\l3}\big)^2\,\Psi_{imnj}(t)\nn\\
  &\quad\qquad-\big(Y_{mjn}^{\l1}+Y_{mjn}^{\l3}\big)
  \big(Y_{mjn}^{\l1}+Y_{mjn}^{\l2}\big)\Psi_{imjn}(t)\Big]\nn\\
  &\qquad+(\Omega_{mjn}^{\l}+\epsilon_i)P_{imjn}^{\l}(t),\\
  \ii\frac{d}{dt}Q_{imjn}^{\l}(t)&=\bar{f}_i
  \Big[\big(Y_{ijn}^{\l1}+Y_{ijn}^{\l2}\big)^2\,\Psi_{imnj}(t)\nn\\
  &\quad\qquad-\big(Y_{ijn}^{\l1}+Y_{ijn}^{\l3}\big)
  \big(Y_{ijn}^{\l1}+Y_{ijn}^{\l2}\big)\Psi_{imjn}(t)\Big]\nn\\
  &\qquad+(\Omega_{ijn}^{\l}+\epsilon_m)Q_{imjn}^{\l}(t).
\end{align}
\end{subequations}
These equations together with Eq.~(\ref{eomrho}) form a closed system of ODEs which define
the Faddeev approximation within the GKBA framework.  We emphasize that to obtain
$\Omega^{\l}_{mjn}$ and $Y^{\l}_{mjn}$ we simply have to diagonalize $3\times 3$ matrices
for every $m\in \mathcal{V}$ and for every pair $j,\,n\in \mathcal{O}$.  The numerical
solution of the Faddeev scheme scales linearly with the propagation time and it is
therefore competitive with the conventional NEGF approaches discussed in
Section~\ref{sec:derivation}.

\section{Photoinduced dynamics in glycine}
\label{sec:gly:III}

As a test model for the investigation of the shake-up processes we consider the Gly I
conformer of the glycine molecule in which an XUV pulse creates a hole in the inner
valence states. Glycine is the simplest natural amino acid with just 15 valence molecular
orbitals. Its nontrivial electronic structure~\cite{myhre_x-ray_2019} represents a tough
test for numerical methods as discussed below. The system has been previously studied in a
number of works. Kuleff \emph{et al.}~\cite{kuleff_multielectron_2005,kuleff_charge_2007}
describe in details the periodic charge migration of a hole following its sudden creation
in the $11a'$ MO. They demonstrate that oscillations with period of about 8\,fs between
the $11a'$ and $12a'$ MOs are responsible for the major part of the dynamics. However,
this is also accompanied by the excitation-deexcitation of the $4a''$ and $16a'$ MOs, and
by the promotion of an electron to the unoccupied $5a''$ MO. This picture was confirmed
using the NEGF-2B method~\cite{perfetto_first-principles_2019}. Very similar quantum
beatings between $11a'$ and $12a'$ have been predicted in
Ref.~\cite{cooper_single-photon_2013}; here the authors also propose a mechanism to
experimentally detect the effect using the so-called single-photon laser-enabled Auger
decay. Finally, we mention a recent DFT study tuned towards more realististic description
of the initial photoinoization~\cite{ayuso_ultrafast_2017}\,---\,the attosecond XUV pulse
is explicitly taken into account leading to the broad 17 to 35\,eV spectrum of
excitations.

\begin{table}[t]
  \caption{\label{HFenes} HF energies of the five MOs participating to the reduced
    photoinduced dynamics of the glycine molecule. The energy level positions are
    indicated according to the aufbau principle with respect to highest occupied molecular
    orbital (HOMO) and lowest unoccupied molecular orbital (LUMO).}
  \renewcommand{\arraystretch}{1.2}
  \begin{ruledtabular}
    \begin{tabular}{ccd{2}@{\hspace{0.8cm}}}
      \text{State}& Position &\multicolumn{1}{c}{HF~Energy~(eV)}\\
      \hline
      $11a'$ &HOMO-9 & -19.15 \\
      $12a'$ &HOMO-8 & -18.74 \\
      $4a''$ &HOMO-2 & -12.93 \\
      $16a'$ &HOMO   & -10.86 \\
      $5a''$ &LUMO+3 & + 4.79 \\
       \end{tabular}
    \end{ruledtabular}
\end{table}

We consider here a reduced Hamiltonian for the Gly I conformer which takes into account
only the five HF MOs involved in the dynamics of the $8$~fs charge oscillation, namely the
occupied states $11a'$, $12a'$, $4a''$ and $16a'$ and one unoccupied state $5a''$.  The
occupancies of all other valence states is frozen to the equilibrium value.  The HF
energies of the relevant MOs is reported in Table~\ref{HFenes}.  We refer to our previous
works on the electronic structure of the ground and excited states, basis representation,
and femtosecond dynamics of this
molecule~\cite{perfetto_first-principles_2019,perfetto_cheers:_2018}.  The reduced system
is ionized by coupling the MOs to a fictitious vacuum state through
$\Omega(t)=E(t)\mathcal{D}$ where $\mathcal{D}$ is the dipole matrix element (chosen
independent of the states) and $E(t)$ is the electric field of a weak attosecond XUV pulse
causing single-photon ionizations.  To better highlight correlation effects we did not
consider pulse-induced transitions between different MOs, see below.  We perform our
calculations at the fixed geometry since the nuclear dynamics is expected to take place at
longer time-scales. However, this is an important
ingredient~\cite{li_ultrafast_2015,lara-astiaso_role_2017,polyak_charge_2018} to make
theory predictive in experimental energy- and time-ranges.

\begin{figure*}
\includegraphics[width=1\textwidth]{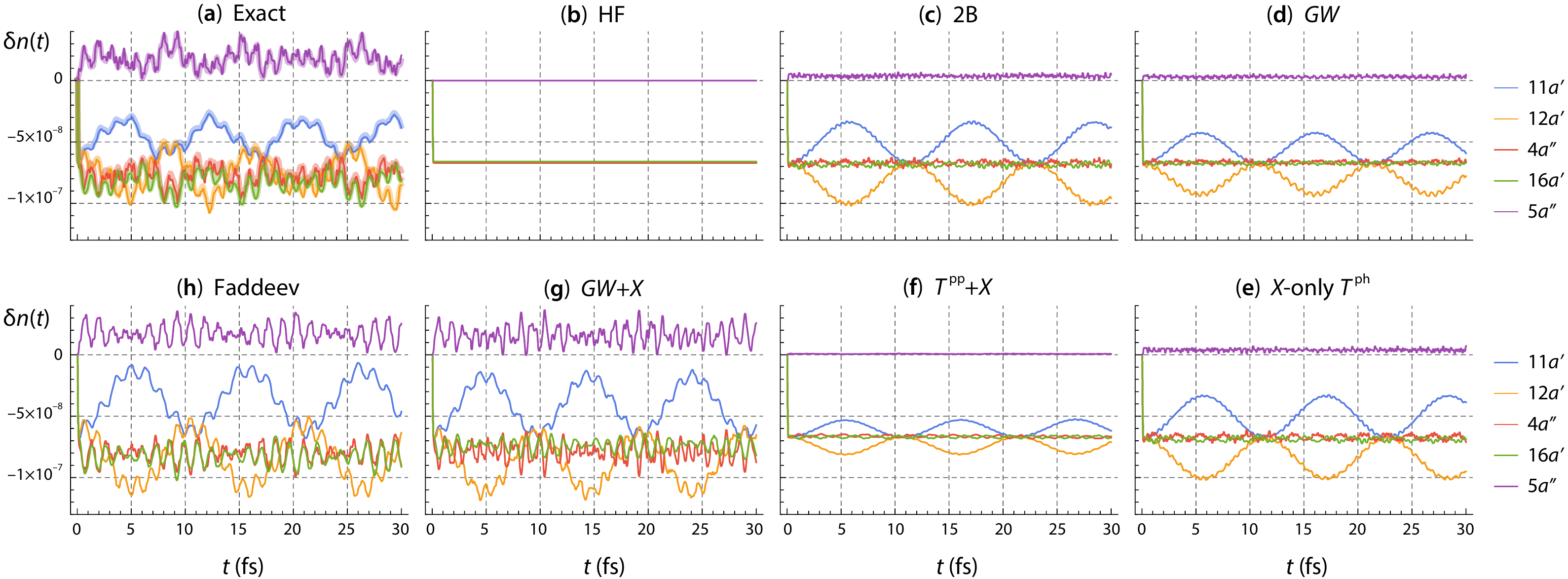}\\
\caption{Electron occupancies of the five MO of the glycine 
molecule after photoionization. (a) Exact solution of the
Schr{\"o}dinger equation in the subspace spanned by the 5 MOs and 
the fictitious vacuum state (thin lines). Additionally we demonstrate that the dynamics 
can be resonably represented by truncating the CI expansion to 
$2h$-$1p$ states (thick lines). (b)-(h) GKBA simulations in different 
approximations.}
\label{fig-occupations}
\end{figure*}

In Fig.~\ref{fig-occupations}\,(a) we show the time-dependent change of the MO occupancies
as obtained from the exact solution of the Schr{\"o}dinger equation in the subspace of the
5+1 states (thin lines). Additionally we demonstrate that the dynamics can be resonably
represented by taking into account only $1h$ an $2h$-$1p$ states in the configuration
interaction (CI) expansion (thick lines).  This implies that shake-up processes dominate
the correlation-driven dynamics. During the action of the XUV pulse the occupied states
loose charge mainly due to photoionization. Shake-up $1h\rightarrow 2h$-$1p$ processes
initiate immediately after the pulse and are responsible for populating the virtual
(unoccupied) state $5a''$. Time-dependent HF simulations clearly show the crucial role
played by correlations, see panel~(b). The HF Hamiltonian remains essentially the same
after the pulse as only $\simeq 10^{-8}$ electrons are expelled. Since pulse-induced
transitions between MOs have been neglected, the occupancies remain almost constant, and
in particular the virtual state does not populate.

The considered subspace of 5 MOs well capture the 8\,fs oscillation of the $11a'$ and
$12a'$ occupancies, see again panel~(a).  Although this effect can be described in terms
of simple $1h$ transitions between the involved MOs, the HF approximation remains
inadequate. This is due to the fact that the cationic states $\hat{d}_{11a'}|\phi_{\rm
  HF}\ket$ and $\hat{d}_{12a'}|\phi_{\rm HF}\ket$ are not exact eigenstates of $\hat{H}$
(excited state correlations).  As we shall see almost all correlated methods cure this
problem; they are able to describe the bounce of charge between the $11a'$ and $12a'$ MOs,
albeit with slightly different periods. A secondary, yet dominat, feature is the
superimposed oscillation of higher frequency, with a period $\simeq 1.4$~fs.  A careful
inspection reveals that this faster mode can be associated to the $1h\rightarrow 2h$-$1p$
transition
\be
\hat{d}_{12a'}|\phi_{\rm HF}\ket\to \hat{d}^{\dag}_{5a''}\hat{d}_{16a'}\hat{d}_{4a''}|\phi_{\rm HF}\ket.
\label{transition}
\ee
It turns out that this mode is much more difficult to predict, and so far it has not been
accessed by any of the existing methods.

To appreciate the difficulty we have performed 2B, $GW$ and $T$-matrix simulations with
and without exchange ($X$) diagrams.  All these methods bring about some correlations
already in the neutral ground state and, thus, it seems unavoidable to perform the
adiabatic switching procedure in order to construct a {\em stationary} correlated ground
state. As already discussed, however, the initial ground state of glycine is well
approximated by a single Slater determinant and it is therefore accurate to start the
simulation from the HF ground state.  The question then arises, how the adiabatic
switching can be avoided in such a way that the HF ground state is a stationary solution
of the GKBA equation (\ref{eomrho}) in the absence of external fields? The answer to this
question is rooted in the physics of the photoinduced dynamics. The main role of the
collision integral is to initiate the shake-up process. Following the reasoning that has
led us to develop the Faddeev approximation, we replace $v$ with $v^{s}$ in
Eq.~(\ref{eq:Is}), compare with Eq.~(\ref{eq:I:s:0}). Furthermore, the expansion of $\mcG$
to lowest order in $v^{s}$ amounts to replace $v$ with $v^{s}$ also in Eq.~(\ref{Psidef}).
The full Coulomb tensor $\mw$ is instead retained in the products $\mw\mrho^{\Delta}$ and
$\mrho^{\Delta}\mw$ of Eq.~(\ref{eq:G2:X:init}) in order to fully account for the repeated
scattering between particles in the virtual $2h$-$1p$ states.  It is easy to show that
this adjustment is equivalent to calculate the self-energy using $v^{s}$ instead of $v$ in
the {\em external} interaction lines. With this adjustment, the HF density matrix is
stationary in the absence of external field for any correlated method since the $v^{s}$
Coulomb tensor has only one index in the $\mathcal{V}$-sector and $\mPsi$ contains the
product of two $\rho^{<}=\mathrm{diag}\{1,1,1,1,0\}$ and two
$\rho^{>}=\mathrm{diag}\{0,0,0,0,1\}$\,---\,this implies that the driving term at the
initial time, i.\,e., $\mPsi(0)$ vanishes.

In Fig.~\ref{fig-occupations}\,(c) we show the results of the simplest correlated
approximation, i.\,e., the 2B approximation. Due to the lack of $h$-$h$ and $p$-$h$
scatterings, the energy of the $2h$-$1p$ state is simply given by \be \Omega^{\rm 2B}=
\e_{5a''}-\e_{16a'}-\e_{4a''} \ee and hence the transition energy $\Omega^{\rm
  2B}+\e_{12a'}\simeq 9.8$\,eV, corresponding to a period of $0.42$\,fs, is severely
overestimated.  The situation does not improve in the $GW$ approximation, see panel~(d),
nor in the $T$-matrix approximation in the $pp$ channel (almost the same as in $GW$ and
hence not shown). The $T$-matrix approximation in the $ph$ channel is unstable toward the
formation of strongly bound electron-hole pairs; therefore we do not have results to show
for $T^{ph}$. As anticipated the failure of these methods must be attributed to the
absence of $2h$-$1p$ correlations.  As a matter of fact they do not even take into account
virtual spin-flip scatterings in the $p$-$h$ (for $GW$ and $T^{ph}$) or $p$-$p$ (for
$T^{pp}$) channels since exchange diagrams are discarded.

The inclusion of exchange diagrams does not, in general, guarantee a better
performance. In panel~(e) we show the results of a simulation using the $T$-matrix
approximation in the $ph$ channel with only exchange diagrams ($X$-only). Although
$X$-only $T^{ph}$ is stable, the 1.4~fs oscillation is absent. We could perform
simulations with both direct and exchange diagrams for $GW$ and $T^{pp}$. Surprisingly we
found that $GW$+$X$ provides a key improvement, see panel~(g), whereas exchange diagrams
in $T^{pp}$ play essentially no role, see panel~(f). The rational behind these outcomes
should be searched in the values of the direct and exchange Coulomb integrals, i.\,e.,
$v^{d}_{\m\n}$ and $v^{x}_{\m\n}$, responsible for renormalizing the energy of the
$2h$-$1p$ states, see Eqs.~(\ref{eq:h:eff}-\ref{eq:v:2index}).  It turns out that the
$12a'$ hole of spin $\s$ is mainly coupled (through $v^{s}$) to the $2h$-$1p$ states
$16a'_{\s}4a''_{\ua}-5a''_{\da}$ and $16a'_{\s}4a''_{\da}-5a''_{\ua}$, which are in turn
coupled by the anomalously large exchange integral $v^{x}_{4a'',5a''}\simeq 2.3$\,eV (all
other exchange integrals are negligible).  The energy of these two $2h$-$1p$ states is
almost the same and given by
\be
\Omega\simeq \Omega^{\rm 2B}+v^{d}_{4a'',16a'}-v^{d}_{5a'',16a'}-
v^{d}_{4a'',5a''}+v^{x}_{4a'',5a''}.
\label{omegaci}
\ee
The direct integrals are all large with $v^{d}_{4a'',16a'}\simeq v^{d}_{5a'',16a'}\simeq
6.5$\,eV and $v^{d}_{4a'',5a''}\simeq 10.6$\,eV. Due to the cancellation between the first
two direct integrals in Eq.~(\ref{omegaci}) only the direct and exchange integrals with
labels $4a'',5a''$ are relevant in $\Omega$. These are precisely the ones taken into
account by the $GW$ approximation, see discussion below Eq.~(\ref{faddeevGr}).  The
inclusion of exchange, i.\,e., $GW$+$X$, provides a key improvement of the theory since it
describes the spin-flip scattering process mediated by $v^{x}_{4a'',5a''}$. We conclude
that the good performance of the $GW$+$X$ approximation is a {\em mere coincidence} as it
strongly relies on the particular values of the Coulomb integrals in glycine.

Time-dependent simulations in the Faddeev approximation are shown in
Fig.~\ref{fig-occupations}\,(h). The results are of comparable quality to the $GW$+$X$
ones, in agreement with the discussion above. However, the Faddeev approximation does not
rely on any special values of the Coulomb integrals\,---\,$2h$-$1p$ correlations are fully
taken into account. This is reflected in a slightly more accurate value of the period of
the superimposed oscillations, $1.33$\,fs against the $1.2$\,fs in $GW$+$X$ (we recall
that the exact value is $1.4$\,fs).

\begin{figure}[]
  {\includegraphics[width=0.99\columnwidth]{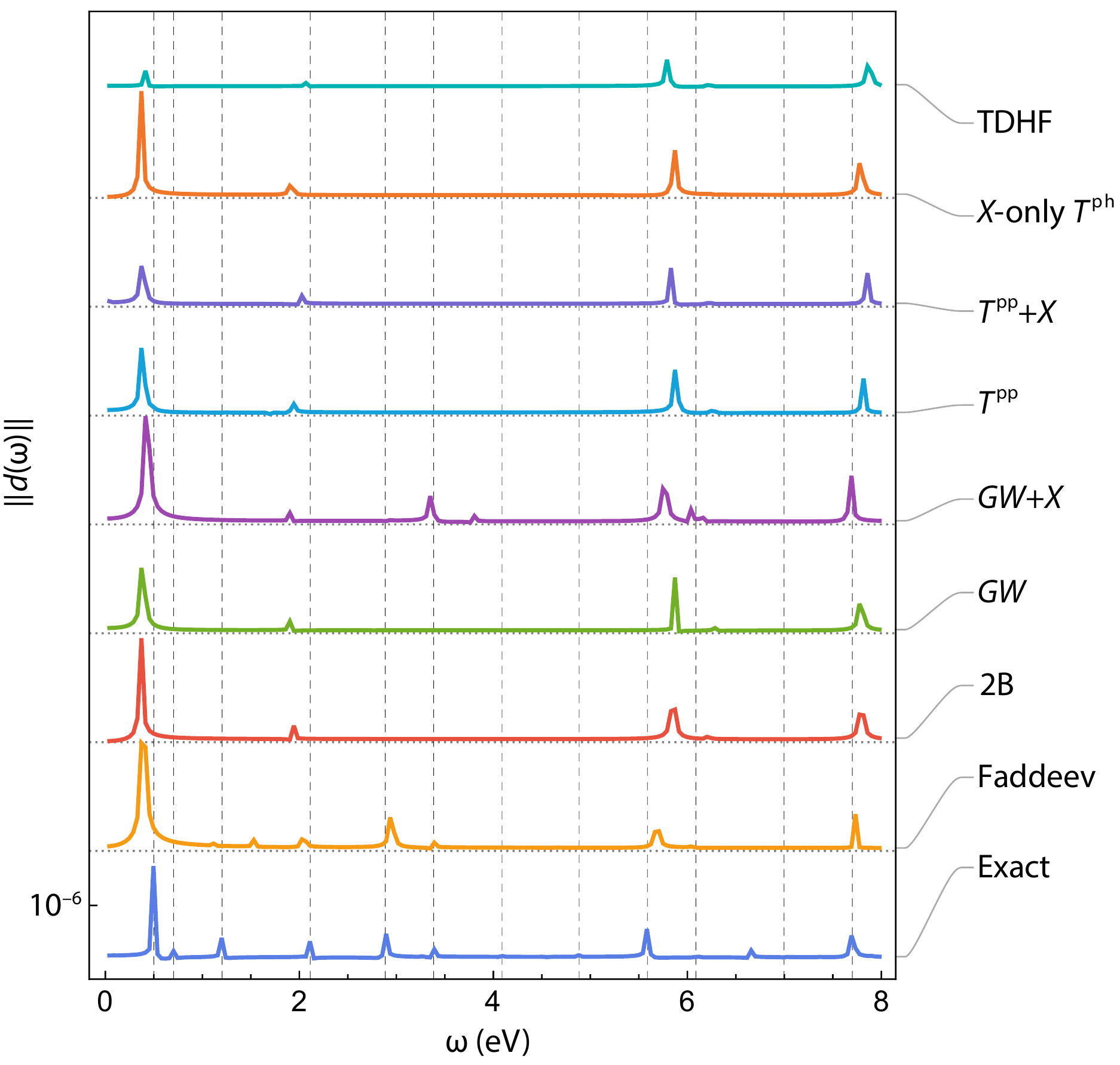}}
  \caption[]{Power spectrum computed as the Fourier transform of the photoinduced dipole
    moment. The peaks associated to the $11a'\leftrightarrow 12a'$ quantum beating at
    energy 0.5\,eV (period $\simeq 8$\,fs) and the shake-up processes involving the
    electron promotion to the unoccupied $5a''$ state at energy $2.9$\,eV (period $\simeq
    1.4$\,fs) are clearly visible. Notice, that the number of peaks is smaller than in the
    DFT analysis of Ayuso \emph{et al.}~\cite{ayuso_ultrafast_2017} because of the minimal
    model considered here.\label{fig:gl-power}}
\end{figure}

The occupations of the MOs coincide with the diagonal elements of the one-particle density
matrix $\rho^{<}$. As the GKBA approach returns the full density matrix we could also
investigate how accurate the off-diagonal elements are.  For this purpose we have
calculated the photoinduced dipole moment
\be
 d_\a(t)=\sum_{ij}d^{\a}_{ij}\rho^{<}_{ji}(t),
\ee
where $d^{\a}_{ij}$ are the dipole matrix elements along the direction $\a$ calculated in
Ref.~\cite{perfetto_first-principles_2019}, and then extracted the power spectrum from the
Fourier transform, $\lVert d(\omega)\rVert^2=\frac{1}{3}\sum_\alpha|d_\a(\omega)|^2$.  The
outcome of exact and GKBA simulations is shown in Fig.~\ref{fig:gl-power}. With the
exception of $GW$+$X$ and Faddev all other approximations yield only four peaks; their
origin is essentially the same as in HF, see top panel, although different approximations
give different weights.  The $GW$+$X$ represents a clear improvement over all other
methods but visible discrepancies occur here too. The lowest energy and higher energy
peaks are well reproduced but all other peaks are either misplaced by hundreds of meV or
completely absent.  In contrast the Faddeev approximation captures with high accuracy all
main peaks except for the second and third low energy ones (whose energy is overstimated)
and the one at energy $\simeq 6.7$~eV which is missing.

\section{Conclusions}
\label{sec:conclusions}
In conclusion, we have provided an accurate NEGF description and an efficient
implementation scheme for the ubiquitous shake-up mechanism which accompanies the
ultrafast valence-hole migration in organic molecules triggered by a weak XUV pulse.
Calculations based on the unifying matrix formalism clearly demonstrate that none of the
state-of-the-art NEGF methods such as second Born, $GW$ or $T$-matrix are capable to
describe it. Our solution has been inspired by the three-particle Faddeev approach which
treats $2h$-$1p$ scatterings non-perturbatively and it relies on an extension of the
original GKBA to higher order Green's functions.  The Faddeev-NEGF scheme scales linearly
in time opening prospects for the incorporation of other effects such as interaction with
collective nuclear and electronic excitations and the inclusion of continuum scattering
states for the accurate description of ultrafast spectroscopies of organic molecules.

\begin{acknowledgments}
  We thank Tommasso Mazzocchi for his help during the early stages of this work. We
  acknowledge the financial support from MIUR PRIN (Grant No. 20173B72NB), from INFN
  through the TIME2QUEST project, and from Tor Vergata University through the Beyond
  Borders Project ULEXIEX.
\end{acknowledgments}

\appendix

\section{Intuition behind GKBA and its generalization to higher-order GFs}
\label{gkbaintroapp}
Let us start by ``deriving'' the generalized Kadanoff-Baym ansatz. This is just an
approximation that can intuitively be derived from the following considerations for the
\emph{mean-field} GF. Let us express our main quantity as
\be
G_0^{<}(t_1,t_2)=U(t_1,t_0)G_0^{<}(t_0,t_0)U(t_0,t_2),\label{eq:g:less:0}
\ee
where $U(t_0,t)$ is the usual time-evolution operator
\be
U(t,t_0)=T\left\{e^{-\ii \int_{t_0}^t d\tau\, h_\text{HF}(\tau)}\right\}.
\label{eq:def:U}
\ee
Eqs.~(\ref{eq:g:less:0},\ref{eq:def:U}) are understood in matrix form.  Using the
semi-group property of the time-evolution operator we split the time-dependence in
Eq.~\eqref{eq:g:less:0}
\begin{multline}
  G_0^{<}(t_1,t_2)=\overbrace{\theta(t_1-t_2)U(t_1,t_2)}\overbrace{U(t_2,t_0)G_0^{<}(t_0,t_0)U(t_0,t_2)}\\
  +\underbrace{U(t_1,t_0)G_0^{<}(t_0,t_0)U(t_0,t_1)}\underbrace{U(t_1,t_2)\theta(t_2-t_1)}.
\end{multline}
Recall now that Hartree-Fock retarded (advanced) GFs fulfill the 
equations of motion (\ref{eq:eom:G:RA}), and therefore they can be written in terms of
the evolution operator
\begin{subequations}
\begin{align}
 G_0^\mathit{R}(t_1,t_2)&=-\ii\theta(t_1-t_2)U(t_1,t_2),\\
 G_0^\mathit{A}(t_1,t_2)&=+ \ii\theta(t_2-t_1)U(t_1,t_2),
\end{align}
\label{eq:GRA:0}%
\end{subequations}
allowing us to re-write
\begin{align}
G_0^{<}(t_1,t_2)=\ii G_0^\mathit{R}(t_1,t_2)G_0^{<}(t_2,t_2)
-\ii G_0^{<}(t_1,t_1)G_0^\mathit{A}(t_1,t_2).
\label{eq:gkba2}
\end{align}
Analogous considerations hold for the greater GF. Now the crutial step is to perform the
replacements $G_0^{\lessgtr}(t_1,t_2)\rightarrow G^{\lessgtr}(t_1,t_2)$ and
$G_0^{\lessgtr}(t,t)\rightarrow \ii \rho^{\lessgtr}(t)$ because the main point of GKBA is
to approximate the \emph{interacting} correlators. This approximation is physically
justified provided that, e.\,g., the quasiparticle life-time is greater than the averaged
electron collision time~\cite{lipavsky_generalized_1986}, and it leads us to the following
compact form
\begin{align}
  G^{\lessgtr}(t_1,t_2)&=-G^\mathit{R}(t_1,t_2)\rho^{\lessgtr}(t_2)
  +\rho^{\lessgtr}(t_1)G^\mathit{A}(t_1,t_2).
  \label{eq:gkba3}
\end{align}

Equation~\eqref{eq:gkba3} allows for further generalizations in the case of more
complicated \emph{two-times} correlators. Consider for instance a very general greater
correlator
\begin{align}
  \mathcal{G}^{>}(1,2)=\frac{1}{\ii^n}\langle\hat{\mathcal{C}}_H(\bar{x}_1,t_1)
  \hat{\mathcal{C}}_H^\dagger(\bar{x}_2,t_2)\rangle,
  \label{eq:GOO:def}
\end{align}
where $1\equiv(\bar{x}_1,t_1)$, etc., for brevity, $\hat{\mathcal{C}}_H(\bar{x},t)$ being
a \emph{composite} operator that can be expressed as a product of $n$ fermionic creation
$\hat{d}^\dagger$ and annihilation $\hat{d}$ operators in the Heisenberg picture, and
$\bar{x}$ being a \emph{collective coordinate} associated with the product.  Our goal is
to devise a GKBA for the correlator~\eqref{eq:GOO:def} starting again with a correlator
averaged over $|\phi_\text{HF}\rangle$. In order to simplify the discussion, we introduce
a new set of fermionic operators $\hat{c}$ and $\hat{c}^\dagger$ as to make the
Hartree-Fock state $|\phi_\text{HF}\rangle$ to be the vaccuum state, which we will denote
for brevity as $|\vac\rangle$. Specifically we have
\begin{align}
\hat{c}_i
&=\begin{cases}
    \hat{d}_i & i\in \mathcal{O}\\
    \hat{d}_i^\dagger & i\in \mathcal{V}
\end{cases},&
\hat{c}_i^\dagger
&=\begin{cases}
    \hat{d}_i^\dagger & i\in \mathcal{V}\\
    \hat{d}_i & i\in \mathcal{O}
\end{cases},
\end{align}
where $\mathcal{O}$ denotes the set of occupied states and $\mathcal{V}$ is the set of
unoccupied states. With these definitions
\be
\hat c_i|\vac\rangle=0,
\label{eq:vac}
\ee
and the only operators for which the mean-field approximation to the
correlator~\eqref{eq:GOO:def}
\begin{align}
  \mathcal{G}_0^{>}(\bar{x}_1,t_1;\bar{x}_2,t_2)=\frac{1}{\ii^n}\langle\vac(t_1)|\hat{\mathcal{C}}_{\bar{x}_1}\hat{U}(t_1,t_2)
  \hat{\mathcal{C}}_{\bar{x}_2}^\dagger|\vac(t_2)\rangle
  \label{eq:c2}
\end{align}
is nonvanishing are those given by the product
\be
\hat{\mathcal{C}}_{\bar{x}_1}=\hat{c}_{\bar{x}_{11}}\hat{c}_{\bar{x}_{12}}\ldots\hat{c}_{\bar{x}_{1n}}.
\ee
As we mention above, this convenience is one of the reasons of introducing new fermionic
operators.

In Eq.~\eqref{eq:c2}, we expanded the operators in the Heisenberg picture, introduced the
time-evolution operator $\hat{U}(t_1,t_2)$, and embedded some of the time-dependence into
the bra- and ket-states.  Consider now the states
\[
|\bar{y}\rangle=\hat{\mathcal{C}}_{\bar{y}}^\dagger|\vac\rangle,
\]
which form a complete orthonormal system. The completeness relation
\be
\frac{1}{n!}\sum_{\bar{y}}|\bar{y}\rangle\langle\bar{y}|=\mathbb{1}
\ee
can be used in order to factorize $\mathcal{G}_0^{>}$. There is a certain freedom on where
it can be inserted. In order to build parallels with Eq.~\eqref{eq:gkba2} we split
Eq.~\eqref{eq:c2} into two parts, proportional to $\theta(t_1-t_2)$ and $\theta(t_2-t_1)$,
respectively. In the first part, the completness relation is inserted after
$\hat{U}(t_1,t_2)$, and in the second part\,---\,before it. As the consequence we obtain a
generalization of Eq.~\eqref{eq:gkba2}
\begin{multline}  
  \langle\vac(t_1)|\hat{\mathcal{C}}_{\bar{x}_1}\hat{U}(t_1,t_2)
  \hat{\mathcal{C}}_{\bar{x}_2}^\dagger|\vac(t_2)\rangle\\
  =\frac{1}{n!}\theta(t_1-t_2)\sum_{\bar{y}}
  \langle\vac(t_1)|\hat{\mathcal{C}}_{\bar{x}_1}\hat{U}(t_1,t_2)\hat{\mathcal{C}}_{\bar{y}}^\dagger|\vac(t_2)\rangle\\
  \times\langle\vac(t_2)|\hat{\mathcal{C}}_{\bar{y}}\hat{\mathcal{C}}_{\bar{x}_2}^\dagger|\vac(t_2)\rangle
  +\frac{1}{n!}\theta(t_2-t_1)\sum_{\bar{y}}
  \langle\vac(t_1)|\hat{\mathcal{C}}_{\bar{x}_1}\hat{\mathcal{C}}_{\bar{y}}^\dagger|\vac(t_1)\rangle\\
  \times\langle\vac(t_1)|\hat{\mathcal{C}}_{\bar{y}}\hat{U}(t_1,t_2)\hat{\mathcal{C}}_{\bar{x}_2}^\dagger|\vac(t_2)\rangle.
  \label{eq:fct}
\end{multline}
Let us introduce the retarded and advanced correlators
\begin{subequations}
  \label{eq:GRA:def}
\begin{align}
  \mathcal{G}^\mathit{R}(t_1,t_2)&=-\frac{\ii}{n!}\theta(t_1-t_2)
  \big\langle\big[\hat{\mathcal{C}}_H(\bar{x}_1,t_1),\hat{\mathcal{C}}_H^\dagger(\bar{x}_2,t_2)\big]\big\rangle,\\
  \mathcal{G}^\mathit{A}(t_1,t_2)&=+ \frac{\ii}{n!}\theta(t_2-t_1)
  \big\langle \big[\hat{\mathcal{C}}_H(\bar{x}_1,t_1),\hat{\mathcal{C}}_H^\dagger(\bar{x}_2,t_2)\big]\big\rangle.
\end{align}
\end{subequations}
This form is chosen to put them in correspondence with the $n$-body time-evolution
operators, cf. Eq.~\eqref{eq:GRA:0}. We furthermore notice the presence of equal-time
correlators in Eq.~\eqref{eq:fct} such as
$\langle\vac(t_2)|\hat{\mathcal{C}}_{\bar{y}}\hat{\mathcal{C}}_{\bar{x}_2}^\dagger|\vac(t_2)\rangle$
and
$\langle\vac(t_1)|\hat{\mathcal{C}}_{\bar{x}_1}\hat{\mathcal{C}}_{\bar{y}}^\dagger|\vac(t_1)\rangle$. They
are analogous to the single-particle densities in Eq.~\eqref{eq:gkba3}. Performing now a
transition to the correlated reference state in Eq.~\eqref{eq:fct}, using definitions
Eqs.~(\ref{eq:GOO:def} and \ref{eq:GRA:def}), and considering that the same arguments
apply to the lesser correlator, we finally obtain
\begin{multline}
  \mathcal{G}^{\lessgtr}(\bar{x}_1,t_1;\bar{x}_2,t_2)=\ii \sum_{\bar{y}}\mathcal{G}^\mathit{R}(\bar{x}_1,t_1;\bar{y},t_2)
  \mathcal{G}^{\lessgtr}(\bar{y},t_2;\bar{x}_2,t_2)\\
  -\ii\sum_{\bar{y}}\mathcal{G}^{\lessgtr}(\bar{x}_1,t_1;\bar{y},t_1)\mathcal{G}^\mathit{A}(\bar{y},t_1;\bar{x}_2,t_2).
  \label{eq:GKBA:gen}
\end{multline}
Notice that in order to introduce the retarded and the advanced GFs in these equations we used
\begin{multline}
\frac{1}{n!}\theta(t_1-t_2)\langle\vac| \hat{\mathcal{C}}_H(\bar{x}_1,t_1),\hat{\mathcal{C}}_H^\dagger(\bar{x}_2,t_2)|\vac\rangle\\
=\frac{1}{n!}\theta(t_1-t_2)\big\langle\vac|\big[ \hat{\mathcal{C}}_H(\bar{x}_1,t_1),\hat{\mathcal{C}}_H^\dagger(\bar{x}_2,t_2)\big]|\vac\big\rangle\\
=\ii\mathcal{G}_0^\mathit{R}(t_1,t_2),
\end{multline}
where the commutator can be introduced in view of the special choice of operators
(Eq.~\ref{eq:vac}) that guarantee that $\hat{\mathcal{C}}_{\bar{x}_1}|\phi\rangle=0$.

At first glance, Eq.~\eqref{eq:GKBA:gen} seems to be just a trivial generalization of GKBA
to many-particle scenarios. However, let us inspect the physical content of even simpler
$\mathcal{G}_0^\mathit{R/A}(1,2)$ correlators. They are computed with the ordinary HF
Hamiltonian, however, on the subspace of $n$-particle excitations making it similar to the
multiconfiguration time-dependent Hartree-Fock
approach~\cite{szalay_multiconfiguration_2011}. This represents a completely novel aspect
of our theory. We remind the reader that in Sec.~\ref{sec:spin} we have $n=3$, i.\,e., with
the help of GKBA~\eqref{eq:GKBA:gen}, we factorize the $2h$-$1p$ GF~\eqref{eq:G2h1p:def}
into a product of two terms: the one that contains 3-particle spin correlations and the
other one that contains the population dynamics, viz. Eq.~\eqref{eq:G2h1p:fct}. In order
to obtain this equation we explicitly set
\begin{subequations}
\begin{align}
  \hat{\mathcal{C}}^\dagger_{\bar{x}_1}&
  =\hat{d}_{n\s_{3}}^\dagger \hat{d}_{j\s_{2}}^\dagger \hat{d}_{m\s_{1}},&
  \hat{\mathcal{C}}_{\bar{y}}&
  =\hat{d}_{m\s''_{1}}^\dagger \hat{d}_{j\s''_{2}}^\dagger \hat{d}_{n\s''_{1}},\\
  \hat{\mathcal{C}}_{\bar{y}}^\dagger&
  =\hat{d}_{n\s''_{3}}^\dagger \hat{d}_{j\s''_{2}}^\dagger \hat{d}_{m\s''_{1}},&
  \hat{\mathcal{C}}_{\bar{x}_2}&
  =\hat{d}_{p\s'_{1}}^\dagger \hat{d}_{r\s'_{2}} \hat{d}_{s\s'_{3}}.
\end{align}
\end{subequations}
As can be seen from the definition of $\hat{\mathcal{C}}_{\bar{y}}$, we exploit the
factorization of the many-body states only in the spin-sector. The equal-time $2h$-$1p$
correlators in Eq.~\eqref{eq:G2h1p:fct} are further computed with the help of the Wick's
theorem:
\begin{multline*}
  \langle\hat{\mathcal{C}}_{\bar{y}}^\dagger \hat{\mathcal{C}}_{\bar{x}_2}\rangle
  =\langle\hat{d}_{n\s''_{3}}^\dagger \hat{d}_{j\s''_{2}}^\dagger \hat{d}_{m\s''_{1}}
  \hat{d}_{p\s'_{1}}^\dagger \hat{d}_{r\s'_{2}} \hat{d}_{s\s'_{3}}
  \rangle\\
  =\delta_{\s''_{1}\s'_{1}}\rho_{mp}^>\left\{\delta_{\s''_{2}\s'_{2}}\delta_{\s''_{3}\s'_{3}} \rho_{rj}^<\rho_{sn}^<
  -\delta_{\s''_{2}\s'_{3}}\delta_{\s''_{3}\s'_{2}} \rho_{rn}^<\rho_{sj}^<
  \right\}.
\end{multline*}  
\section{Some nonequilibrium identities}
\label{sec:app:id}
According to the Langreth rules~\cite{stefanucci_nonequilibrium_2013} we have
\begin{align}
  \mchi^{<}=\mchi^{0,<}+\mchi^R\cdot\mw\cdot\mchi^{0,<}+\mchi^<\cdot\mw\cdot\mchi^{0,A},
  \label{eq:app:1}
\end{align}
where quite generally the $R/A$-components are defined in terms of the $\lessgtr$-components
\begin{align*}
  A^{R}(t,t')&=+\theta(t-t')\big\{A^>(t,\,t')-A^<(t,\,t')\big\},\\
  A^{A}(t,t')&=-\theta(t'-t)\big\{A^>(t,\,t')-A^<(t,\,t')\big\}.
\end{align*}
Regrouping the terms in Eq.~\eqref{eq:app:1}, we obtain
\begin{align}
  \mchi^{<}\cdot(\md-\mw\cdot\mchi^{0,A})
  =(\md+\mchi^R\cdot\mw)\cdot\mchi^{0,<}.
  \label{eq:app:2}
\end{align}
Now realize with the help of RPA
\begin{multline*}
 (\md-\mv\cdot\mchi^{0,A})\cdot(\md+\mw\cdot\mchi^{A})
 =\md \\
 -\mw\cdot\big[\mchi^{0,A}-\mchi^{A}\big]
 -\mw\cdot\underbrace{\big[\mchi^{0,A}\cdot\mw\cdot\mchi^{A}\big]}_{=\big[\mchi^{A}-\mchi^{0,A}\big]}
    =\md.
\end{multline*}
Using this identity in Eq.~\ref{eq:app:2} we obtain
\begin{align}
  \mchi^{<}
  =(\md+\mchi^R\cdot\mw)\cdot\mchi^{0,<}\cdot(\md+\mw\cdot\mchi^{A}).
\end{align}

\begin{thebibliography}{67}%
\makeatletter
\providecommand \@ifxundefined [1]{%
 \@ifx{#1\undefined}
}%
\providecommand \@ifnum [1]{%
 \ifnum #1\expandafter \@firstoftwo
 \else \expandafter \@secondoftwo
 \fi
}%
\providecommand \@ifx [1]{%
 \ifx #1\expandafter \@firstoftwo
 \else \expandafter \@secondoftwo
 \fi
}%
\providecommand \natexlab [1]{#1}%
\providecommand \enquote  [1]{``#1''}%
\providecommand \bibnamefont  [1]{#1}%
\providecommand \bibfnamefont [1]{#1}%
\providecommand \citenamefont [1]{#1}%
\providecommand \href@noop [0]{\@secondoftwo}%
\providecommand \href [0]{\begingroup \@sanitize@url \@href}%
\providecommand \@href[1]{\@@startlink{#1}\@@href}%
\providecommand \@@href[1]{\endgroup#1\@@endlink}%
\providecommand \@sanitize@url [0]{\catcode `\\12\catcode `\$12\catcode
  `\&12\catcode `\#12\catcode `\^12\catcode `\_12\catcode `\%12\relax}%
\providecommand \@@startlink[1]{}%
\providecommand \@@endlink[0]{}%
\providecommand \url  [0]{\begingroup\@sanitize@url \@url }%
\providecommand \@url [1]{\endgroup\@href {#1}{\urlprefix }}%
\providecommand \urlprefix  [0]{URL }%
\providecommand \Eprint [0]{\href }%
\providecommand \doibase [0]{http://dx.doi.org/}%
\providecommand \selectlanguage [0]{\@gobble}%
\providecommand \bibinfo  [0]{\@secondoftwo}%
\providecommand \bibfield  [0]{\@secondoftwo}%
\providecommand \translation [1]{[#1]}%
\providecommand \BibitemOpen [0]{}%
\providecommand \bibitemStop [0]{}%
\providecommand \bibitemNoStop [0]{.\EOS\space}%
\providecommand \EOS [0]{\spacefactor3000\relax}%
\providecommand \BibitemShut  [1]{\csname bibitem#1\endcsname}%
\let\auto@bib@innerbib\@empty
\bibitem [{\citenamefont {Zhang}\ and\ \citenamefont
  {Averitt}(2014)}]{zhang_dynamics_2014}%
  \BibitemOpen
  \bibfield  {author} {\bibinfo {author} {\bibfnamefont {J.}~\bibnamefont
  {Zhang}}\ and\ \bibinfo {author} {\bibfnamefont {R.}~\bibnamefont
  {Averitt}},\ }\href@noop {} {\bibfield  {journal} {\bibinfo  {journal} {Annu.
  Rev. Mater. Res.}\ }\textbf {\bibinfo {volume} {44}},\ \bibinfo {pages} {19}
  (\bibinfo {year} {2014})}\BibitemShut {NoStop}%
\bibitem [{\citenamefont {L\'{e}pine}\ \emph {et~al.}(2014)\citenamefont
  {L\'{e}pine}, \citenamefont {Ivanov},\ and\ \citenamefont
  {Vrakking}}]{lepine_attosecond_2014}%
  \BibitemOpen
  \bibfield  {author} {\bibinfo {author} {\bibfnamefont {F.}~\bibnamefont
  {L\'{e}pine}}, \bibinfo {author} {\bibfnamefont {M.~Y.}\ \bibnamefont
  {Ivanov}}, \ and\ \bibinfo {author} {\bibfnamefont {M.~J.~J.}\ \bibnamefont
  {Vrakking}},\ }\href@noop {} {\bibfield  {journal} {\bibinfo  {journal} {Nat.
  Photonics}\ }\textbf {\bibinfo {volume} {8}},\ \bibinfo {pages} {195}
  (\bibinfo {year} {2014})}\BibitemShut {NoStop}%
\bibitem [{\citenamefont {Kraus}\ \emph {et~al.}(2018)\citenamefont {Kraus},
  \citenamefont {Z\"{u}rch}, \citenamefont {Cushing}, \citenamefont {Neumark},\
  and\ \citenamefont {Leone}}]{kraus_ultrafast_2018}%
  \BibitemOpen
  \bibfield  {author} {\bibinfo {author} {\bibfnamefont {P.~M.}\ \bibnamefont
  {Kraus}}, \bibinfo {author} {\bibfnamefont {M.}~\bibnamefont {Z\"{u}rch}},
  \bibinfo {author} {\bibfnamefont {S.~K.}\ \bibnamefont {Cushing}}, \bibinfo
  {author} {\bibfnamefont {D.~M.}\ \bibnamefont {Neumark}}, \ and\ \bibinfo
  {author} {\bibfnamefont {S.~R.}\ \bibnamefont {Leone}},\ }\href@noop {}
  {\bibfield  {journal} {\bibinfo  {journal} {Nature Reviews Chemistry}\
  }\textbf {\bibinfo {volume} {2}},\ \bibinfo {pages} {82} (\bibinfo {year}
  {2018})}\BibitemShut {NoStop}%
\bibitem [{\citenamefont {Calegari}\ \emph {et~al.}(2014)\citenamefont
  {Calegari}, \citenamefont {Ayuso}, \citenamefont {Trabattoni}, \citenamefont
  {Belshaw}, \citenamefont {De~Camillis}, \citenamefont {Anumula},
  \citenamefont {Frassetto}, \citenamefont {Poletto}, \citenamefont {Palacios},
  \citenamefont {Decleva}, \citenamefont {Greenwood}, \citenamefont {Martin},\
  and\ \citenamefont {Nisoli}}]{calegari_ultrafast_2014}%
  \BibitemOpen
  \bibfield  {author} {\bibinfo {author} {\bibfnamefont {F.}~\bibnamefont
  {Calegari}}, \bibinfo {author} {\bibfnamefont {D.}~\bibnamefont {Ayuso}},
  \bibinfo {author} {\bibfnamefont {A.}~\bibnamefont {Trabattoni}}, \bibinfo
  {author} {\bibfnamefont {L.}~\bibnamefont {Belshaw}}, \bibinfo {author}
  {\bibfnamefont {S.}~\bibnamefont {De~Camillis}}, \bibinfo {author}
  {\bibfnamefont {S.}~\bibnamefont {Anumula}}, \bibinfo {author} {\bibfnamefont
  {F.}~\bibnamefont {Frassetto}}, \bibinfo {author} {\bibfnamefont
  {L.}~\bibnamefont {Poletto}}, \bibinfo {author} {\bibfnamefont
  {A.}~\bibnamefont {Palacios}}, \bibinfo {author} {\bibfnamefont
  {P.}~\bibnamefont {Decleva}}, \bibinfo {author} {\bibfnamefont {J.~B.}\
  \bibnamefont {Greenwood}}, \bibinfo {author} {\bibfnamefont {F.}~\bibnamefont
  {Martin}}, \ and\ \bibinfo {author} {\bibfnamefont {M.}~\bibnamefont
  {Nisoli}},\ }\href@noop {} {\bibfield  {journal} {\bibinfo  {journal}
  {Science}\ }\textbf {\bibinfo {volume} {346}},\ \bibinfo {pages} {336}
  (\bibinfo {year} {2014})}\BibitemShut {NoStop}%
\bibitem [{\citenamefont {Iablonskyi}\ \emph {et~al.}(2017)\citenamefont
  {Iablonskyi}, \citenamefont {Ueda}, \citenamefont {Ishikawa}, \citenamefont
  {Kheifets}, \citenamefont {Carpeggiani}, \citenamefont {Reduzzi},
  \citenamefont {Ahmadi}, \citenamefont {Comby}, \citenamefont {Sansone},
  \citenamefont {Csizmadia}, \citenamefont {Kuehn}, \citenamefont {Ovcharenko},
  \citenamefont {Mazza}, \citenamefont {Meyer}, \citenamefont {Fischer},
  \citenamefont {Callegari}, \citenamefont {Plekan}, \citenamefont {Finetti},
  \citenamefont {Allaria}, \citenamefont {Ferrari}, \citenamefont {Roussel},
  \citenamefont {Gauthier}, \citenamefont {Giannessi},\ and\ \citenamefont
  {Prince}}]{iablonskyi_observation_2017}%
  \BibitemOpen
  \bibfield  {author} {\bibinfo {author} {\bibfnamefont {D.}~\bibnamefont
  {Iablonskyi}}, \bibinfo {author} {\bibfnamefont {K.}~\bibnamefont {Ueda}},
  \bibinfo {author} {\bibfnamefont {K.~L.}~\bibnamefont {Ishikawa}}, \bibinfo
  {author} {\bibfnamefont {A.~S.}~\bibnamefont {Kheifets}}, \bibinfo {author}
  {\bibfnamefont {P.}~\bibnamefont {Carpeggiani}}, \bibinfo {author}
  {\bibfnamefont {M.}~\bibnamefont {Reduzzi}}, \bibinfo {author} {\bibfnamefont
  {H.}~\bibnamefont {Ahmadi}}, \bibinfo {author} {\bibfnamefont
  {A.}~\bibnamefont {Comby}}, \bibinfo {author} {\bibfnamefont
  {G.}~\bibnamefont {Sansone}}, \bibinfo {author} {\bibfnamefont
  {T.}~\bibnamefont {Csizmadia}}, \bibinfo {author} {\bibfnamefont
  {S.}~\bibnamefont {Kuehn}}, \bibinfo {author} {\bibfnamefont
  {E.}~\bibnamefont {Ovcharenko}}, \bibinfo {author} {\bibfnamefont
  {T.}~\bibnamefont {Mazza}}, \bibinfo {author} {\bibfnamefont
  {M.}~\bibnamefont {Meyer}}, \bibinfo {author} {\bibfnamefont
  {A.}~\bibnamefont {Fischer}}, \bibinfo {author} {\bibfnamefont
  {C.}~\bibnamefont {Callegari}}, \bibinfo {author} {\bibfnamefont
  {O.}~\bibnamefont {Plekan}}, \bibinfo {author} {\bibfnamefont
  {P.}~\bibnamefont {Finetti}}, \bibinfo {author} {\bibfnamefont
  {E.}~\bibnamefont {Allaria}}, \bibinfo {author} {\bibfnamefont
  {E.}~\bibnamefont {Ferrari}}, \bibinfo {author} {\bibfnamefont
  {E.}~\bibnamefont {Roussel}}, \bibinfo {author} {\bibfnamefont
  {D.}~\bibnamefont {Gauthier}}, \bibinfo {author} {\bibfnamefont
  {L.}~\bibnamefont {Giannessi}}, \ and\ \bibinfo {author} {\bibfnamefont
  {K.~C.}~\bibnamefont {Prince}},\ }\href@noop {} {\bibfield  {journal} {\bibinfo
  {journal} {Phys. Rev. Lett.}\ }\textbf {\bibinfo {volume} {119}},\ \bibinfo
  {pages} {073203} (\bibinfo {year} {2017})}\BibitemShut {NoStop}%
\bibitem [{\citenamefont {Lara-Astiaso}\ \emph {et~al.}(2018)\citenamefont
  {Lara-Astiaso}, \citenamefont {Galli}, \citenamefont {Trabattoni},
  \citenamefont {Palacios}, \citenamefont {Ayuso}, \citenamefont {Frassetto},
  \citenamefont {Poletto}, \citenamefont {De~Camillis}, \citenamefont
  {Greenwood}, \citenamefont {Decleva}, \citenamefont {Tavernelli},
  \citenamefont {Calegari}, \citenamefont {Nisoli},\ and\ \citenamefont
  {Mart\'{i}n}}]{lara-astiaso_attosecond_2018}%
  \BibitemOpen
  \bibfield  {author} {\bibinfo {author} {\bibfnamefont {M.}~\bibnamefont
  {Lara-Astiaso}}, \bibinfo {author} {\bibfnamefont {M.}~\bibnamefont {Galli}},
  \bibinfo {author} {\bibfnamefont {A.}~\bibnamefont {Trabattoni}}, \bibinfo
  {author} {\bibfnamefont {A.}~\bibnamefont {Palacios}}, \bibinfo {author}
  {\bibfnamefont {D.}~\bibnamefont {Ayuso}}, \bibinfo {author} {\bibfnamefont
  {F.}~\bibnamefont {Frassetto}}, \bibinfo {author} {\bibfnamefont
  {L.}~\bibnamefont {Poletto}}, \bibinfo {author} {\bibfnamefont
  {S.}~\bibnamefont {De~Camillis}}, \bibinfo {author} {\bibfnamefont
  {J.}~\bibnamefont {Greenwood}}, \bibinfo {author} {\bibfnamefont
  {P.}~\bibnamefont {Decleva}}, \bibinfo {author} {\bibfnamefont
  {I.}~\bibnamefont {Tavernelli}}, \bibinfo {author} {\bibfnamefont
  {F.}~\bibnamefont {Calegari}}, \bibinfo {author} {\bibfnamefont
  {M.}~\bibnamefont {Nisoli}}, \ and\ \bibinfo {author} {\bibfnamefont
  {F.}~\bibnamefont {Mart\'{i}n}},\ }\href@noop {} {\bibfield  {journal}
  {\bibinfo  {journal} {J. Phys. Chem. Lett.}\ }\textbf {\bibinfo {volume}
  {9}},\ \bibinfo {pages} {4570} (\bibinfo {year} {2018})}\BibitemShut
  {NoStop}%
\bibitem [{\citenamefont {Herv\'{e}}\ \emph {et~al.}(2020)\citenamefont
  {Herv\'{e}}, \citenamefont {Despr\'{e}}, \citenamefont {Castellanos~Nash},
  \citenamefont {Loriot}, \citenamefont {Boyer}, \citenamefont {Scognamiglio},
  \citenamefont {Karras}, \citenamefont {Br\'{e}dy}, \citenamefont {Constant},
  \citenamefont {Tielens}, \citenamefont {Kuleff},\ and\ \citenamefont
  {L\'{e}pine}}]{herve_ultrafast_2020}%
  \BibitemOpen
  \bibfield  {author} {\bibinfo {author} {\bibfnamefont {M.}~\bibnamefont
  {Herv\'{e}}}, \bibinfo {author} {\bibfnamefont {V.}~\bibnamefont
  {Despr\'{e}}}, \bibinfo {author} {\bibfnamefont {P.}~\bibnamefont
  {Castellanos~Nash}}, \bibinfo {author} {\bibfnamefont {V.}~\bibnamefont
  {Loriot}}, \bibinfo {author} {\bibfnamefont {A.}~\bibnamefont {Boyer}},
  \bibinfo {author} {\bibfnamefont {A.}~\bibnamefont {Scognamiglio}}, \bibinfo
  {author} {\bibfnamefont {G.}~\bibnamefont {Karras}}, \bibinfo {author}
  {\bibfnamefont {R.}~\bibnamefont {Br\'{e}dy}}, \bibinfo {author}
  {\bibfnamefont {E.}~\bibnamefont {Constant}}, \bibinfo {author}
  {\bibfnamefont {A.~G. G.~M.}\ \bibnamefont {Tielens}}, \bibinfo {author}
  {\bibfnamefont {A.~I.}\ \bibnamefont {Kuleff}}, \ and\ \bibinfo {author}
  {\bibfnamefont {F.}~\bibnamefont {L\'{e}pine}},\ }\href@noop {} {\bibfield
  {journal} {\bibinfo  {journal} {Nature Phys.}\ } (\bibinfo {year}
  {2020})}\BibitemShut {NoStop}%
\bibitem [{\citenamefont {Kuleff}\ \emph {et~al.}(2005)\citenamefont {Kuleff},
  \citenamefont {Breidbach},\ and\ \citenamefont
  {Cederbaum}}]{kuleff_multielectron_2005}%
  \BibitemOpen
  \bibfield  {author} {\bibinfo {author} {\bibfnamefont {A.~I.}\ \bibnamefont
  {Kuleff}}, \bibinfo {author} {\bibfnamefont {J.}~\bibnamefont {Breidbach}}, \
  and\ \bibinfo {author} {\bibfnamefont {L.~S.}\ \bibnamefont {Cederbaum}},\
  }\href@noop {} {\bibfield  {journal} {\bibinfo  {journal} {J. Chem. Phys.}\
  }\textbf {\bibinfo {volume} {123}},\ \bibinfo {pages} {044111} (\bibinfo
  {year} {2005})}\BibitemShut {NoStop}%
\bibitem [{\citenamefont {Szalay}\ \emph {et~al.}(2011)\citenamefont {Szalay},
  \citenamefont {M\"{u}ller}, \citenamefont {Gidofalvi}, \citenamefont
  {Lischka},\ and\ \citenamefont {Shepard}}]{szalay_multiconfiguration_2011}%
  \BibitemOpen
  \bibfield  {author} {\bibinfo {author} {\bibfnamefont {P.~G.}\ \bibnamefont
  {Szalay}}, \bibinfo {author} {\bibfnamefont {T.}~\bibnamefont {M\"{u}ller}},
  \bibinfo {author} {\bibfnamefont {G.}~\bibnamefont {Gidofalvi}}, \bibinfo
  {author} {\bibfnamefont {H.}~\bibnamefont {Lischka}}, \ and\ \bibinfo
  {author} {\bibfnamefont {R.}~\bibnamefont {Shepard}},\ }\href@noop {}
  {\bibfield  {journal} {\bibinfo  {journal} {Chem. Rev.}\ }\textbf {\bibinfo
  {volume} {112}},\ \bibinfo {pages} {108} (\bibinfo {year}
  {2011})}\BibitemShut {NoStop}%
\bibitem [{\citenamefont {Popova-Gorelova}\ \emph {et~al.}(2016)\citenamefont
  {Popova-Gorelova}, \citenamefont {K\"{u}pper},\ and\ \citenamefont
  {Santra}}]{popova-gorelova_imaging_2016}%
  \BibitemOpen
  \bibfield  {author} {\bibinfo {author} {\bibfnamefont {D.}~\bibnamefont
  {Popova-Gorelova}}, \bibinfo {author} {\bibfnamefont {J.}~\bibnamefont
  {K\"{u}pper}}, \ and\ \bibinfo {author} {\bibfnamefont {R.}~\bibnamefont
  {Santra}},\ }\href@noop {} {\bibfield  {journal} {\bibinfo  {journal} {Phys.
  Rev. A}\ }\textbf {\bibinfo {volume} {94}},\ \bibinfo {pages} {013412}
  (\bibinfo {year} {2016})}\BibitemShut {NoStop}%
\bibitem [{\citenamefont {Sch\"{u}ler}\ \emph {et~al.}(2016)\citenamefont
  {Sch\"{u}ler}, \citenamefont {Pavlyukh}, \citenamefont {Bolognesi},
  \citenamefont {Avaldi},\ and\ \citenamefont
  {Berakdar}}]{schuler_electron_2016}%
  \BibitemOpen
  \bibfield  {author} {\bibinfo {author} {\bibfnamefont {M.}~\bibnamefont
  {Sch\"{u}ler}}, \bibinfo {author} {\bibfnamefont {Y.}~\bibnamefont
  {Pavlyukh}}, \bibinfo {author} {\bibfnamefont {P.}~\bibnamefont {Bolognesi}},
  \bibinfo {author} {\bibfnamefont {L.}~\bibnamefont {Avaldi}}, \ and\ \bibinfo
  {author} {\bibfnamefont {J.}~\bibnamefont {Berakdar}},\ }\href@noop {}
  {\bibfield  {journal} {\bibinfo  {journal} {Sci. Rep.}\ }\textbf {\bibinfo
  {volume} {6}},\ \bibinfo {pages} {24396} (\bibinfo {year}
  {2016})}\BibitemShut {NoStop}%
\bibitem [{\citenamefont {Usenko}\ \emph {et~al.}(2016)\citenamefont {Usenko},
  \citenamefont {Sch\"{u}ler}, \citenamefont {Azima}, \citenamefont {Jakob},
  \citenamefont {Lazzarino}, \citenamefont {Pavlyukh}, \citenamefont
  {Przystawik}, \citenamefont {Drescher}, \citenamefont {Laarmann},\ and\
  \citenamefont {Berakdar}}]{usenko_femtosecond_2016}%
  \BibitemOpen
  \bibfield  {author} {\bibinfo {author} {\bibfnamefont {S.}~\bibnamefont
  {Usenko}}, \bibinfo {author} {\bibfnamefont {M.}~\bibnamefont {Sch\"{u}ler}},
  \bibinfo {author} {\bibfnamefont {A.}~\bibnamefont {Azima}}, \bibinfo
  {author} {\bibfnamefont {M.}~\bibnamefont {Jakob}}, \bibinfo {author}
  {\bibfnamefont {L.~L.}\ \bibnamefont {Lazzarino}}, \bibinfo {author}
  {\bibfnamefont {Y.}~\bibnamefont {Pavlyukh}}, \bibinfo {author}
  {\bibfnamefont {A.}~\bibnamefont {Przystawik}}, \bibinfo {author}
  {\bibfnamefont {M.}~\bibnamefont {Drescher}}, \bibinfo {author}
  {\bibfnamefont {T.}~\bibnamefont {Laarmann}}, \ and\ \bibinfo {author}
  {\bibfnamefont {J.}~\bibnamefont {Berakdar}},\ }\href@noop {} {\bibfield
  {journal} {\bibinfo  {journal} {New J. Phys.}\ }\textbf {\bibinfo {volume}
  {18}},\ \bibinfo {pages} {113055} (\bibinfo {year} {2016})}\BibitemShut
  {NoStop}%
\bibitem [{\citenamefont {Cuniberti}\ \emph {et~al.}(2005)\citenamefont
  {Cuniberti}, \citenamefont {Fagas},\ and\ \citenamefont
  {Richter}}]{me-book1}%
  \BibitemOpen
  \bibfield  {author} {\bibinfo {author} {\bibfnamefont {G.}~\bibnamefont
  {Cuniberti}}, \bibinfo {author} {\bibfnamefont {G.}~\bibnamefont {Fagas}}, \
  and\ \bibinfo {author} {\bibfnamefont {K.}~\bibnamefont {Richter}},\
  }\href@noop {} {\emph {\bibinfo {title} {Introducing Molecular
  Electronics}}}\ (\bibinfo  {publisher} {Springer},\ \bibinfo {address}
  {Heidelberg},\ \bibinfo {year} {2005})\BibitemShut {NoStop}%
\bibitem [{\citenamefont {Cuevas}\ and\ \citenamefont
  {Scheer}(2010)}]{me-book2}%
  \BibitemOpen
  \bibfield  {author} {\bibinfo {author} {\bibfnamefont {J.}~\bibnamefont
  {Cuevas}}\ and\ \bibinfo {author} {\bibfnamefont {E.}~\bibnamefont
  {Scheer}},\ }\href@noop {} {\emph {\bibinfo {title} {Molecular Electronics:
  An Introduction to Theory and Experiment}}}\ (\bibinfo  {publisher} {World
  Scientific},\ \bibinfo {address} {London},\ \bibinfo {year}
  {2010})\BibitemShut {NoStop}%
\bibitem [{\citenamefont {Cardona}\ and\ \citenamefont
  {Ley}(1978)}]{cardona_photoemission_1978}%
  \BibitemOpen
  \bibinfo {editor} {\bibfnamefont {M.}~\bibnamefont {Cardona}}\ and\ \bibinfo
  {editor} {\bibfnamefont {L.}~\bibnamefont {Ley}},\ eds.,\ \href@noop {}
  {\emph {\bibinfo {title} {Photoemission in {Solids} {I} {General}
  {Principles}}}}\ (\bibinfo  {publisher} {Springer},\ \bibinfo {address}
  {Berlin},\ \bibinfo {year} {1978})\BibitemShut {NoStop}%
\bibitem [{\citenamefont {Freericks}\ \emph {et~al.}(2009)\citenamefont
  {Freericks}, \citenamefont {Krishnamurthy},\ and\ \citenamefont
  {Pruschke}}]{freericks_theoretical_2009}%
  \BibitemOpen
  \bibfield  {author} {\bibinfo {author} {\bibfnamefont {J.~K.}\ \bibnamefont
  {Freericks}}, \bibinfo {author} {\bibfnamefont {H.~R.}\ \bibnamefont
  {Krishnamurthy}}, \ and\ \bibinfo {author} {\bibfnamefont {T.}~\bibnamefont
  {Pruschke}},\ }\href@noop {} {\bibfield  {journal} {\bibinfo  {journal}
  {Phys. Rev. Lett.}\ }\textbf {\bibinfo {volume} {102}},\ \bibinfo {pages}
  {136401} (\bibinfo {year} {2009})}\BibitemShut {NoStop}%
\bibitem [{\citenamefont {Pavlyukh}\ \emph {et~al.}(2015)\citenamefont
  {Pavlyukh}, \citenamefont {Sch\"{u}ler},\ and\ \citenamefont
  {Berakdar}}]{pavlyukh_single-_2015}%
  \BibitemOpen
  \bibfield  {author} {\bibinfo {author} {\bibfnamefont {Y.}~\bibnamefont
  {Pavlyukh}}, \bibinfo {author} {\bibfnamefont {M.}~\bibnamefont
  {Sch\"{u}ler}}, \ and\ \bibinfo {author} {\bibfnamefont {J.}~\bibnamefont
  {Berakdar}},\ }\href@noop {} {\bibfield  {journal} {\bibinfo  {journal}
  {Phys. Rev. B}\ }\textbf {\bibinfo {volume} {91}},\ \bibinfo {pages} {155116}
  (\bibinfo {year} {2015})}\BibitemShut {NoStop}%
\bibitem [{\citenamefont {Ruberti}\ \emph {et~al.}(2014)\citenamefont
  {Ruberti}, \citenamefont {Averbukh},\ and\ \citenamefont
  {Decleva}}]{ruberti_b-spline_2014}%
  \BibitemOpen
  \bibfield  {author} {\bibinfo {author} {\bibfnamefont {M.}~\bibnamefont
  {Ruberti}}, \bibinfo {author} {\bibfnamefont {V.}~\bibnamefont {Averbukh}}, \
  and\ \bibinfo {author} {\bibfnamefont {P.}~\bibnamefont {Decleva}},\
  }\href@noop {} {\bibfield  {journal} {\bibinfo  {journal} {J. Chem. Phys.}\
  }\textbf {\bibinfo {volume} {141}},\ \bibinfo {pages} {164126} (\bibinfo
  {year} {2014})}\BibitemShut {NoStop}%
\bibitem [{\citenamefont {Ruberti}\ \emph
  {et~al.}(2018{\natexlab{a}})\citenamefont {Ruberti}, \citenamefont
  {Decleva},\ and\ \citenamefont {Averbukh}}]{ruberti_full_2018}%
  \BibitemOpen
  \bibfield  {author} {\bibinfo {author} {\bibfnamefont {M.}~\bibnamefont
  {Ruberti}}, \bibinfo {author} {\bibfnamefont {P.}~\bibnamefont {Decleva}}, \
  and\ \bibinfo {author} {\bibfnamefont {V.}~\bibnamefont {Averbukh}},\
  }\href@noop {} {\bibfield  {journal} {\bibinfo  {journal} {J. Chem. Theory
  Comput.}\ }\textbf {\bibinfo {volume} {14}},\ \bibinfo {pages} {4991}
  (\bibinfo {year} {2018}{\natexlab{a}})}\BibitemShut {NoStop}%
\bibitem [{\citenamefont {Ruberti}\ \emph
  {et~al.}(2018{\natexlab{b}})\citenamefont {Ruberti}, \citenamefont
  {Decleva},\ and\ \citenamefont {Averbukh}}]{ruberti_multi-channel_2018}%
  \BibitemOpen
  \bibfield  {author} {\bibinfo {author} {\bibfnamefont {M.}~\bibnamefont
  {Ruberti}}, \bibinfo {author} {\bibfnamefont {P.}~\bibnamefont {Decleva}}, \
  and\ \bibinfo {author} {\bibfnamefont {V.}~\bibnamefont {Averbukh}},\
  }\href@noop {} {\bibfield  {journal} {\bibinfo  {journal} {Phys. Chem. Chem.
  Phys.}\ }\textbf {\bibinfo {volume} {20}},\ \bibinfo {pages} {8311} (\bibinfo
  {year} {2018}{\natexlab{b}})}\BibitemShut {NoStop}%
\bibitem [{\citenamefont {Pathak}\ \emph {et~al.}(2020)\citenamefont {Pathak},
  \citenamefont {Sato},\ and\ \citenamefont
  {Ishikawa}}]{pathak_time-dependent_2020}%
  \BibitemOpen
  \bibfield  {author} {\bibinfo {author} {\bibfnamefont {H.}~\bibnamefont
  {Pathak}}, \bibinfo {author} {\bibfnamefont {T.}~\bibnamefont {Sato}}, \ and\
  \bibinfo {author} {\bibfnamefont {K.~L.}\ \bibnamefont {Ishikawa}},\
  }\href@noop {} {\bibfield  {journal} {\bibinfo  {journal} {J. Chem. Phys.}\
  }\textbf {\bibinfo {volume} {152}},\ \bibinfo {pages} {124115} (\bibinfo
  {year} {2020})}\BibitemShut {NoStop}%
\bibitem [{\citenamefont {Andreussi}\ \emph {et~al.}(2015)\citenamefont
  {Andreussi}, \citenamefont {Knecht}, \citenamefont {Marian}, \citenamefont
  {Kongsted},\ and\ \citenamefont {Mennucci}}]{andreussi_carotenoids_2015}%
  \BibitemOpen
  \bibfield  {author} {\bibinfo {author} {\bibfnamefont {O.}~\bibnamefont
  {Andreussi}}, \bibinfo {author} {\bibfnamefont {S.}~\bibnamefont {Knecht}},
  \bibinfo {author} {\bibfnamefont {C.~M.}\ \bibnamefont {Marian}}, \bibinfo
  {author} {\bibfnamefont {J.}~\bibnamefont {Kongsted}}, \ and\ \bibinfo
  {author} {\bibfnamefont {B.}~\bibnamefont {Mennucci}},\ }\href@noop {}
  {\bibfield  {journal} {\bibinfo  {journal} {J. Chem. Theory Comput.}\
  }\textbf {\bibinfo {volume} {11}},\ \bibinfo {pages} {655} (\bibinfo {year}
  {2015})}\BibitemShut {NoStop}%
\bibitem [{\citenamefont {Nisoli}\ \emph {et~al.}(2017)\citenamefont {Nisoli},
  \citenamefont {Decleva}, \citenamefont {Calegari}, \citenamefont {Palacios},\
  and\ \citenamefont {Mart\'{i}n}}]{nisoli_attosecond_2017}%
  \BibitemOpen
  \bibfield  {author} {\bibinfo {author} {\bibfnamefont {M.}~\bibnamefont
  {Nisoli}}, \bibinfo {author} {\bibfnamefont {P.}~\bibnamefont {Decleva}},
  \bibinfo {author} {\bibfnamefont {F.}~\bibnamefont {Calegari}}, \bibinfo
  {author} {\bibfnamefont {A.}~\bibnamefont {Palacios}}, \ and\ \bibinfo
  {author} {\bibfnamefont {F.}~\bibnamefont {Mart\'{i}n}},\ }\href@noop {}
  {\bibfield  {journal} {\bibinfo  {journal} {Chem. Rev.}\ }\textbf {\bibinfo
  {volume} {117}},\ \bibinfo {pages} {10760} (\bibinfo {year}
  {2017})}\BibitemShut {NoStop}%
\bibitem [{\citenamefont {Stefanucci}\ and\ \citenamefont {van
  Leeuwen}(2013)}]{stefanucci_nonequilibrium_2013}%
  \BibitemOpen
  \bibfield  {author} {\bibinfo {author} {\bibfnamefont {G.}~\bibnamefont
  {Stefanucci}}\ and\ \bibinfo {author} {\bibfnamefont {R.}~\bibnamefont {van
  Leeuwen}},\ }\href@noop {} {\emph {\bibinfo {title} {Nonequilibrium
  {Many}-{Body} {Theory} of {Quantum} {Systems}: {A} {Modern}
  {Introduction}}}}\ (\bibinfo  {publisher} {Cambridge University Press},\
  \bibinfo {address} {Cambridge},\ \bibinfo {year} {2013})\BibitemShut
  {NoStop}%
\bibitem [{\citenamefont {Balzer}(2013)}]{balzer_nonequilibrium_2013}%
  \BibitemOpen
  \bibfield  {author} {\bibinfo {author} {\bibfnamefont {K.}~\bibnamefont
  {Balzer}},\ }\href@noop {} {\emph {\bibinfo {title} {Nonequilibrium green's
  functions approach to inhomogeneous systems}}},\ \bibinfo {edition} {1st}\
  ed.,\ \bibinfo {series} {Lecture notes in physics}\ No.\ \bibinfo {number}
  {867}\ (\bibinfo  {publisher} {Springer},\ \bibinfo {address} {New York},\
  \bibinfo {year} {2013})\BibitemShut {NoStop}%
\bibitem [{\citenamefont {Lipavsk{\'y}}\ \emph {et~al.}(1986)\citenamefont
  {Lipavsk{\'y}}, \citenamefont {\v{S}pi\v{c}ka},\ and\ \citenamefont
  {Velick{\'y}}}]{lipavsky_generalized_1986}%
  \BibitemOpen
  \bibfield  {author} {\bibinfo {author} {\bibfnamefont {P.}~\bibnamefont
  {Lipavsk{\'y}}}, \bibinfo {author} {\bibfnamefont {V.}~\bibnamefont
  {\v{S}pi\v{c}ka}}, \ and\ \bibinfo {author} {\bibfnamefont {B.}~\bibnamefont
  {Velick{\'y}}},\ }\href@noop {} {\bibfield  {journal} {\bibinfo  {journal}
  {Phys. Rev. B}\ }\textbf {\bibinfo {volume} {34}},\ \bibinfo {pages} {6933}
  (\bibinfo {year} {1986})}\BibitemShut {NoStop}%
\bibitem [{\citenamefont {Karlsson}\ \emph {et~al.}(2018)\citenamefont
  {Karlsson}, \citenamefont {van Leeuwen}, \citenamefont {Perfetto},\ and\
  \citenamefont {Stefanucci}}]{karlsson_generalized_2018}%
  \BibitemOpen
  \bibfield  {author} {\bibinfo {author} {\bibfnamefont {D.}~\bibnamefont
  {Karlsson}}, \bibinfo {author} {\bibfnamefont {R.}~\bibnamefont {van
  Leeuwen}}, \bibinfo {author} {\bibfnamefont {E.}~\bibnamefont {Perfetto}}, \
  and\ \bibinfo {author} {\bibfnamefont {G.}~\bibnamefont {Stefanucci}},\
  }\href@noop {} {\bibfield  {journal} {\bibinfo  {journal} {Phys. Rev. B}\
  }\textbf {\bibinfo {volume} {98}},\ \bibinfo {pages} {115148} (\bibinfo
  {year} {2018})}\BibitemShut {NoStop}%
\bibitem [{\citenamefont {Hermanns}\ \emph {et~al.}(2014)\citenamefont
  {Hermanns}, \citenamefont {Schl\"{u}nzen},\ and\ \citenamefont
  {Bonitz}}]{hermanns_hubbard_2014}%
  \BibitemOpen
  \bibfield  {author} {\bibinfo {author} {\bibfnamefont {S.}~\bibnamefont
  {Hermanns}}, \bibinfo {author} {\bibfnamefont {N.}~\bibnamefont
  {Schl\"{u}nzen}}, \ and\ \bibinfo {author} {\bibfnamefont {M.}~\bibnamefont
  {Bonitz}},\ }\href@noop {} {\bibfield  {journal} {\bibinfo  {journal} {Phys.
  Rev. B}\ }\textbf {\bibinfo {volume} {90}},\ \bibinfo {pages} {125111}
  (\bibinfo {year} {2014})}\BibitemShut {NoStop}%
\bibitem [{\citenamefont {Schl\"{u}nzen}\ \emph {et~al.}(2016)\citenamefont
  {Schl\"{u}nzen}, \citenamefont {Hermanns}, \citenamefont {Bonitz},\ and\
  \citenamefont {Verdozzi}}]{schlunzen_dynamics_2016}%
  \BibitemOpen
  \bibfield  {author} {\bibinfo {author} {\bibfnamefont {N.}~\bibnamefont
  {Schl\"{u}nzen}}, \bibinfo {author} {\bibfnamefont {S.}~\bibnamefont
  {Hermanns}}, \bibinfo {author} {\bibfnamefont {M.}~\bibnamefont {Bonitz}}, \
  and\ \bibinfo {author} {\bibfnamefont {C.}~\bibnamefont {Verdozzi}},\
  }\href@noop {} {\bibfield  {journal} {\bibinfo  {journal} {Phys. Rev. B}\
  }\textbf {\bibinfo {volume} {93}},\ \bibinfo {pages} {035107} (\bibinfo
  {year} {2016})}\BibitemShut {NoStop}%
\bibitem [{\citenamefont {Bar~Lev}\ and\ \citenamefont
  {Reichman}(2014)}]{bar_lev_dynamics_2014}%
  \BibitemOpen
  \bibfield  {author} {\bibinfo {author} {\bibfnamefont {Y.}~\bibnamefont
  {Bar~Lev}}\ and\ \bibinfo {author} {\bibfnamefont {D.~R.}\ \bibnamefont
  {Reichman}},\ }\href@noop {} {\bibfield  {journal} {\bibinfo  {journal}
  {Phys. Rev. B}\ }\textbf {\bibinfo {volume} {89}},\ \bibinfo {pages} {220201(R)}
  (\bibinfo {year} {2014})}\BibitemShut {NoStop}%
\bibitem [{\citenamefont {Latini}\ \emph {et~al.}(2014)\citenamefont {Latini},
  \citenamefont {Perfetto}, \citenamefont {Uimonen}, \citenamefont {van
  Leeuwen},\ and\ \citenamefont {Stefanucci}}]{latini_charge_2014}%
  \BibitemOpen
  \bibfield  {author} {\bibinfo {author} {\bibfnamefont {S.}~\bibnamefont
  {Latini}}, \bibinfo {author} {\bibfnamefont {E.}~\bibnamefont {Perfetto}},
  \bibinfo {author} {\bibfnamefont {A.-M.}\ \bibnamefont {Uimonen}}, \bibinfo
  {author} {\bibfnamefont {R.}~\bibnamefont {van Leeuwen}}, \ and\ \bibinfo
  {author} {\bibfnamefont {G.}~\bibnamefont {Stefanucci}},\ }\href@noop {}
  {\bibfield  {journal} {\bibinfo  {journal} {Phys. Rev. B}\ }\textbf {\bibinfo
  {volume} {89}},\ \bibinfo {pages} {075306} (\bibinfo {year}
  {2014})}\BibitemShut {NoStop}%
\bibitem [{\citenamefont {Cosco}\ \emph {et~al.}(2020)\citenamefont {Cosco},
  \citenamefont {Talarico}, \citenamefont {Tuovinen},\ and\ \citenamefont
  {Gullo}}]{cosco2020spectral}%
  \BibitemOpen
  \bibfield  {author} {\bibinfo {author} {\bibfnamefont {F.}~\bibnamefont
  {Cosco}}, \bibinfo {author} {\bibfnamefont {N.~W.}\ \bibnamefont {Talarico}},
  \bibinfo {author} {\bibfnamefont {R.}~\bibnamefont {Tuovinen}}, \ and\
  \bibinfo {author} {\bibfnamefont {N.~L.}\ \bibnamefont {Gullo}},\ }\href@noop
  {} {} (\bibinfo {year} {2020}),\ \Eprint {http://arxiv.org/abs/2007.08901}
  {arXiv:2007.08901 [cond-mat.str-el]} \BibitemShut {NoStop}%
\bibitem [{\citenamefont {Tuovinen}\ \emph {et~al.}(2021)\citenamefont
  {Tuovinen}, \citenamefont {van Leeuwen}, \citenamefont {Perfetto},\ and\
  \citenamefont {Stefanucci}}]{Riku2021}%
  \BibitemOpen
  \bibfield  {author} {\bibinfo {author} {\bibfnamefont {R.}~\bibnamefont
  {Tuovinen}}, \bibinfo {author} {\bibfnamefont {R.}~\bibnamefont {van
  Leeuwen}}, \bibinfo {author} {\bibfnamefont {E.}~\bibnamefont {Perfetto}}, \
  and\ \bibinfo {author} {\bibfnamefont {G.}~\bibnamefont {Stefanucci}},\
  }\href@noop {} {\bibfield  {journal} {\bibinfo  {journal} {The Journal of
  Chemical Physics}\ }\textbf {\bibinfo {volume} {154}},\ \bibinfo {pages}
  {094104} (\bibinfo {year} {2021})}\BibitemShut {NoStop}%
\bibitem [{\citenamefont {Covito}\ \emph {et~al.}(2018)\citenamefont {Covito},
  \citenamefont {Perfetto}, \citenamefont {Rubio},\ and\ \citenamefont
  {Stefanucci}}]{covito_real-time_2018}%
  \BibitemOpen
  \bibfield  {author} {\bibinfo {author} {\bibfnamefont {F.}~\bibnamefont
  {Covito}}, \bibinfo {author} {\bibfnamefont {E.}~\bibnamefont {Perfetto}},
  \bibinfo {author} {\bibfnamefont {A.}~\bibnamefont {Rubio}}, \ and\ \bibinfo
  {author} {\bibfnamefont {G.}~\bibnamefont {Stefanucci}},\ }\href@noop {}
  {\bibfield  {journal} {\bibinfo  {journal} {Phys. Rev. A}\ }\textbf {\bibinfo
  {volume} {97}},\ \bibinfo {pages} {061401(R)} (\bibinfo {year}
  {2018})}\BibitemShut {NoStop}%
\bibitem [{\citenamefont {Tuovinen}\ \emph {et~al.}(2020)\citenamefont
  {Tuovinen}, \citenamefont {Gole\v{z}}, \citenamefont {Eckstein},\ and\
  \citenamefont {Sentef}}]{tuovinen_comparing_2020}%
  \BibitemOpen
  \bibfield  {author} {\bibinfo {author} {\bibfnamefont {R.}~\bibnamefont
  {Tuovinen}}, \bibinfo {author} {\bibfnamefont {D.}~\bibnamefont {Gole\v{z}}},
  \bibinfo {author} {\bibfnamefont {M.}~\bibnamefont {Eckstein}}, \ and\
  \bibinfo {author} {\bibfnamefont {M.~A.}\ \bibnamefont {Sentef}},\
  }\href@noop {} {\bibfield  {journal} {\bibinfo  {journal} {Phys. Rev. B}\
  }\textbf {\bibinfo {volume} {102}},\ \bibinfo {pages} {115157} (\bibinfo
  {year} {2020})}\BibitemShut {NoStop}%
\bibitem [{\citenamefont {Pal}\ \emph {et~al.}(2011)\citenamefont {Pal},
  \citenamefont {Pavlyukh}, \citenamefont {H\"{u}bner},\ and\ \citenamefont
  {Schneider}}]{pal_optical_2011}%
  \BibitemOpen
  \bibfield  {author} {\bibinfo {author} {\bibfnamefont {G.}~\bibnamefont
  {Pal}}, \bibinfo {author} {\bibfnamefont {Y.}~\bibnamefont {Pavlyukh}},
  \bibinfo {author} {\bibfnamefont {W.}~\bibnamefont {H\"{u}bner}}, \ and\
  \bibinfo {author} {\bibfnamefont {H.~C.}\ \bibnamefont {Schneider}},\
  }\href@noop {} {\bibfield  {journal} {\bibinfo  {journal} {Eur. Phys. J. B}\
  }\textbf {\bibinfo {volume} {79}},\ \bibinfo {pages} {327} (\bibinfo {year}
  {2011})}\BibitemShut {NoStop}%
\bibitem [{\citenamefont {Perfetto}\ \emph
  {et~al.}(2015{\natexlab{a}})\citenamefont {Perfetto}, \citenamefont
  {Uimonen}, \citenamefont {van Leeuwen},\ and\ \citenamefont
  {Stefanucci}}]{perfetto_first-principles_2015}%
  \BibitemOpen
  \bibfield  {author} {\bibinfo {author} {\bibfnamefont {E.}~\bibnamefont
  {Perfetto}}, \bibinfo {author} {\bibfnamefont {A.-M.}\ \bibnamefont
  {Uimonen}}, \bibinfo {author} {\bibfnamefont {R.}~\bibnamefont {van
  Leeuwen}}, \ and\ \bibinfo {author} {\bibfnamefont {G.}~\bibnamefont
  {Stefanucci}},\ }\href@noop {} {\bibfield  {journal} {\bibinfo  {journal}
  {Phys. Rev. A}\ }\textbf {\bibinfo {volume} {92}},\ \bibinfo {pages} {033419}
  (\bibinfo {year} {2015}{\natexlab{a}})}\BibitemShut {NoStop}%
\bibitem [{\citenamefont {Perfetto}\ \emph
  {et~al.}(2015{\natexlab{b}})\citenamefont {Perfetto}, \citenamefont
  {Sangalli}, \citenamefont {Marini},\ and\ \citenamefont
  {Stefanucci}}]{perfetto_nonequilibrium_2015}%
  \BibitemOpen
  \bibfield  {author} {\bibinfo {author} {\bibfnamefont {E.}~\bibnamefont
  {Perfetto}}, \bibinfo {author} {\bibfnamefont {D.}~\bibnamefont {Sangalli}},
  \bibinfo {author} {\bibfnamefont {A.}~\bibnamefont {Marini}}, \ and\ \bibinfo
  {author} {\bibfnamefont {G.}~\bibnamefont {Stefanucci}},\ }\href@noop {}
  {\bibfield  {journal} {\bibinfo  {journal} {Phys. Rev. B}\ }\textbf {\bibinfo
  {volume} {92}},\ \bibinfo {pages} {205304} (\bibinfo {year}
  {2015}{\natexlab{b}})}\BibitemShut {NoStop}%
\bibitem [{\citenamefont {Sangalli}\ \emph {et~al.}(2016)\citenamefont
  {Sangalli}, \citenamefont {Dal~Conte}, \citenamefont {Manzoni}, \citenamefont
  {Cerullo},\ and\ \citenamefont {Marini}}]{sangalli_nonequilibrium_2016}%
  \BibitemOpen
  \bibfield  {author} {\bibinfo {author} {\bibfnamefont {D.}~\bibnamefont
  {Sangalli}}, \bibinfo {author} {\bibfnamefont {S.}~\bibnamefont {Dal~Conte}},
  \bibinfo {author} {\bibfnamefont {C.}~\bibnamefont {Manzoni}}, \bibinfo
  {author} {\bibfnamefont {G.}~\bibnamefont {Cerullo}}, \ and\ \bibinfo
  {author} {\bibfnamefont {A.}~\bibnamefont {Marini}},\ }\href@noop {}
  {\bibfield  {journal} {\bibinfo  {journal} {Phys. Rev. B}\ }\textbf {\bibinfo
  {volume} {93}},\ \bibinfo {pages} {195205} (\bibinfo {year}
  {2016})}\BibitemShut {NoStop}%
\bibitem [{\citenamefont {Pogna}\ \emph {et~al.}(2016)\citenamefont {Pogna},
  \citenamefont {Marsili}, \citenamefont {De~Fazio}, \citenamefont {Dal~Conte},
  \citenamefont {Manzoni}, \citenamefont {Sangalli}, \citenamefont {Yoon},
  \citenamefont {Lombardo}, \citenamefont {Ferrari}, \citenamefont {Marini},
  \citenamefont {Cerullo},\ and\ \citenamefont
  {Prezzi}}]{pogna_photo-induced_2016}%
  \BibitemOpen
  \bibfield  {author} {\bibinfo {author} {\bibfnamefont {E.~A.~A.}\
  \bibnamefont {Pogna}}, \bibinfo {author} {\bibfnamefont {M.}~\bibnamefont
  {Marsili}}, \bibinfo {author} {\bibfnamefont {D.}~\bibnamefont {De~Fazio}},
  \bibinfo {author} {\bibfnamefont {S.}~\bibnamefont {Dal~Conte}}, \bibinfo
  {author} {\bibfnamefont {C.}~\bibnamefont {Manzoni}}, \bibinfo {author}
  {\bibfnamefont {D.}~\bibnamefont {Sangalli}}, \bibinfo {author}
  {\bibfnamefont {D.}~\bibnamefont {Yoon}}, \bibinfo {author} {\bibfnamefont
  {A.}~\bibnamefont {Lombardo}}, \bibinfo {author} {\bibfnamefont {A.~C.}\
  \bibnamefont {Ferrari}}, \bibinfo {author} {\bibfnamefont {A.}~\bibnamefont
  {Marini}}, \bibinfo {author} {\bibfnamefont {G.}~\bibnamefont {Cerullo}}, \
  and\ \bibinfo {author} {\bibfnamefont {D.}~\bibnamefont {Prezzi}},\
  }\href@noop {} {\bibfield  {journal} {\bibinfo  {journal} {ACS Nano}\
  }\textbf {\bibinfo {volume} {10}},\ \bibinfo {pages} {1182} (\bibinfo {year}
  {2016})}\BibitemShut {NoStop}%
\bibitem [{\citenamefont {Sangalli}\ and\ \citenamefont
  {Marini}(2015)}]{sangalli_ultra-fast_2015}%
  \BibitemOpen
  \bibfield  {author} {\bibinfo {author} {\bibfnamefont {D.}~\bibnamefont
  {Sangalli}}\ and\ \bibinfo {author} {\bibfnamefont {A.}~\bibnamefont
  {Marini}},\ }\href@noop {} {\bibfield  {journal} {\bibinfo  {journal}
  {Eurphys. Lett.}\ }\textbf {\bibinfo {volume} {110}},\ \bibinfo {pages}
  {47004} (\bibinfo {year} {2015})}\BibitemShut {NoStop}%
\bibitem [{\citenamefont {Perfetto}\ \emph {et~al.}(2016)\citenamefont
  {Perfetto}, \citenamefont {Sangalli}, \citenamefont {Marini},\ and\
  \citenamefont {Stefanucci}}]{perfetto_first-principles_2016}%
  \BibitemOpen
  \bibfield  {author} {\bibinfo {author} {\bibfnamefont {E.}~\bibnamefont
  {Perfetto}}, \bibinfo {author} {\bibfnamefont {D.}~\bibnamefont {Sangalli}},
  \bibinfo {author} {\bibfnamefont {A.}~\bibnamefont {Marini}}, \ and\ \bibinfo
  {author} {\bibfnamefont {G.}~\bibnamefont {Stefanucci}},\ }\href@noop {}
  {\bibfield  {journal} {\bibinfo  {journal} {Phys. Rev. B}\ }\textbf {\bibinfo
  {volume} {94}},\ \bibinfo {pages} {245303} (\bibinfo {year}
  {2016})}\BibitemShut {NoStop}%
\bibitem [{\citenamefont {Joost}\ \emph {et~al.}(2020)\citenamefont {Joost},
  \citenamefont {Schl\"{u}nzen},\ and\ \citenamefont
  {Bonitz}}]{joost_g1-g2_2020}%
  \BibitemOpen
  \bibfield  {author} {\bibinfo {author} {\bibfnamefont {J.-P.}\ \bibnamefont
  {Joost}}, \bibinfo {author} {\bibfnamefont {N.}~\bibnamefont
  {Schl\"{u}nzen}}, \ and\ \bibinfo {author} {\bibfnamefont {M.}~\bibnamefont
  {Bonitz}},\ }\href@noop {} {\bibfield  {journal} {\bibinfo  {journal} {Phys.
  Rev. B}\ }\textbf {\bibinfo {volume} {101}},\ \bibinfo {pages} {245101}
  (\bibinfo {year} {2020})}\BibitemShut {NoStop}%
\bibitem [{\citenamefont {Dahlen}\ and\ \citenamefont {van
  Leeuwen}(2007)}]{dahlen_solving_2007}%
  \BibitemOpen
  \bibfield  {author} {\bibinfo {author} {\bibfnamefont {N.~E.}\ \bibnamefont
  {Dahlen}}\ and\ \bibinfo {author} {\bibfnamefont {R.}~\bibnamefont {van
  Leeuwen}},\ }\href@noop {} {\bibfield  {journal} {\bibinfo  {journal} {Phys.
  Rev. Lett.}\ }\textbf {\bibinfo {volume} {98}},\ \bibinfo {pages} {153004}
  (\bibinfo {year} {2007})}\BibitemShut {NoStop}%
\bibitem [{\citenamefont {My\"{o}h\"{a}nen}\ \emph {et~al.}(2008)\citenamefont
  {My\"{o}h\"{a}nen}, \citenamefont {Stan}, \citenamefont {Stefanucci},\ and\
  \citenamefont {van Leeuwen}}]{myohanen_many-body_2008}%
  \BibitemOpen
  \bibfield  {author} {\bibinfo {author} {\bibfnamefont {P.}~\bibnamefont
  {My\"{o}h\"{a}nen}}, \bibinfo {author} {\bibfnamefont {A.}~\bibnamefont
  {Stan}}, \bibinfo {author} {\bibfnamefont {G.}~\bibnamefont {Stefanucci}}, \
  and\ \bibinfo {author} {\bibfnamefont {R.}~\bibnamefont {van Leeuwen}},\
  }\href@noop {} {\bibfield  {journal} {\bibinfo  {journal} {Eurphys. Lett.}\
  }\textbf {\bibinfo {volume} {84}},\ \bibinfo {pages} {67001} (\bibinfo {year}
  {2008})}\BibitemShut {NoStop}%
\bibitem [{\citenamefont {My\"{o}h\"{a}nen}\ \emph {et~al.}(2009)\citenamefont
  {My\"{o}h\"{a}nen}, \citenamefont {Stan}, \citenamefont {Stefanucci},\ and\
  \citenamefont {van Leeuwen}}]{myohanen_kadanoff-baym_2009}%
  \BibitemOpen
  \bibfield  {author} {\bibinfo {author} {\bibfnamefont {P.}~\bibnamefont
  {My\"{o}h\"{a}nen}}, \bibinfo {author} {\bibfnamefont {A.}~\bibnamefont
  {Stan}}, \bibinfo {author} {\bibfnamefont {G.}~\bibnamefont {Stefanucci}}, \
  and\ \bibinfo {author} {\bibfnamefont {R.}~\bibnamefont {van Leeuwen}},\
  }\href@noop {} {\bibfield  {journal} {\bibinfo  {journal} {Phys. Rev. B}\
  }\textbf {\bibinfo {volume} {80}},\ \bibinfo {pages} {115107} (\bibinfo
  {year} {2009})}\BibitemShut {NoStop}%
\bibitem [{\citenamefont {Puig~von Friesen}\ \emph {et~al.}(2010)\citenamefont
  {Puig~von Friesen}, \citenamefont {Verdozzi},\ and\ \citenamefont
  {Almbladh}}]{puig_von_friesen_kadanoff-baym_2010}%
  \BibitemOpen
  \bibfield  {author} {\bibinfo {author} {\bibfnamefont {M.}~\bibnamefont
  {Puig~von Friesen}}, \bibinfo {author} {\bibfnamefont {C.}~\bibnamefont
  {Verdozzi}}, \ and\ \bibinfo {author} {\bibfnamefont {C.-O.}\ \bibnamefont
  {Almbladh}},\ }\href@noop {} {\bibfield  {journal} {\bibinfo  {journal}
  {Phys. Rev. B}\ }\textbf {\bibinfo {volume} {82}},\ \bibinfo {pages} {155108}
  (\bibinfo {year} {2010})}\BibitemShut {NoStop}%
\bibitem [{\citenamefont {Friesen}\ \emph {et~al.}(2010)\citenamefont
  {Friesen}, \citenamefont {Verdozzi},\ and\ \citenamefont
  {Almbladh}}]{friesen_artificial_2010}%
  \BibitemOpen
  \bibfield  {author} {\bibinfo {author} {\bibfnamefont {M.~P.~v.}\
  \bibnamefont {Friesen}}, \bibinfo {author} {\bibfnamefont {C.}~\bibnamefont
  {Verdozzi}}, \ and\ \bibinfo {author} {\bibfnamefont {C.-O.}\ \bibnamefont
  {Almbladh}},\ }\href@noop {} {\bibfield  {journal} {\bibinfo  {journal} {J.
  Phys. Conf. Ser.}\ }\textbf {\bibinfo {volume} {220}},\ \bibinfo {pages}
  {012016} (\bibinfo {year} {2010})}\BibitemShut {NoStop}%
\bibitem [{\citenamefont {S\"{a}kkinen}\ \emph {et~al.}(2012)\citenamefont
  {S\"{a}kkinen}, \citenamefont {Manninen},\ and\ \citenamefont {van
  Leeuwen}}]{sakkinen_kadanoffbaym_2012}%
  \BibitemOpen
  \bibfield  {author} {\bibinfo {author} {\bibfnamefont {N.}~\bibnamefont
  {S\"{a}kkinen}}, \bibinfo {author} {\bibfnamefont {M.}~\bibnamefont
  {Manninen}}, \ and\ \bibinfo {author} {\bibfnamefont {R.}~\bibnamefont {van
  Leeuwen}},\ }\href@noop {} {\bibfield  {journal} {\bibinfo  {journal} {New J.
  Phys.}\ }\textbf {\bibinfo {volume} {14}},\ \bibinfo {pages} {013032}
  (\bibinfo {year} {2012})}\BibitemShut {NoStop}%
\bibitem [{\citenamefont {Murakami}\ \emph {et~al.}(2020)\citenamefont
  {Murakami}, \citenamefont {Sch\"{u}ler}, \citenamefont {Takayoshi},\ and\
  \citenamefont {Werner}}]{murakami_ultrafast_2020}%
  \BibitemOpen
  \bibfield  {author} {\bibinfo {author} {\bibfnamefont {Y.}~\bibnamefont
  {Murakami}}, \bibinfo {author} {\bibfnamefont {M.}~\bibnamefont
  {Sch\"{u}ler}}, \bibinfo {author} {\bibfnamefont {S.}~\bibnamefont
  {Takayoshi}}, \ and\ \bibinfo {author} {\bibfnamefont {P.}~\bibnamefont
  {Werner}},\ }\href@noop {} {\bibfield  {journal} {\bibinfo  {journal} {Phys.
  Rev. B}\ }\textbf {\bibinfo {volume} {101}},\ \bibinfo {pages} {035203}
  (\bibinfo {year} {2020})}\BibitemShut {NoStop}%
\bibitem [{\citenamefont {Pavlyukh}\ \emph {et~al.}(2013)\citenamefont
  {Pavlyukh}, \citenamefont {Berakdar},\ and\ \citenamefont
  {Rubio}}]{pavlyukh_initial_2013}%
  \BibitemOpen
  \bibfield  {author} {\bibinfo {author} {\bibfnamefont {Y.}~\bibnamefont
  {Pavlyukh}}, \bibinfo {author} {\bibfnamefont {J.}~\bibnamefont {Berakdar}},
  \ and\ \bibinfo {author} {\bibfnamefont {A.}~\bibnamefont {Rubio}},\
  }\href@noop {} {\bibfield  {journal} {\bibinfo  {journal} {Phys. Rev. B}\
  }\textbf {\bibinfo {volume} {87}},\ \bibinfo {pages} {125101} (\bibinfo
  {year} {2013})}\BibitemShut {NoStop}%
\bibitem [{\citenamefont {Barbieri}\ and\ \citenamefont
  {Dickhoff}(2001)}]{barbieri_faddeev_2001}%
  \BibitemOpen
  \bibfield  {author} {\bibinfo {author} {\bibfnamefont {C.}~\bibnamefont
  {Barbieri}}\ and\ \bibinfo {author} {\bibfnamefont {W.~H.}\ \bibnamefont
  {Dickhoff}},\ }\href@noop {} {\bibfield  {journal} {\bibinfo  {journal}
  {Phys. Rev. C}\ }\textbf {\bibinfo {volume} {63}},\ \bibinfo {pages} {034313}
  (\bibinfo {year} {2001})}\BibitemShut {NoStop}%
\bibitem [{\citenamefont {Faddeev}(1961)}]{faddeev_scattering_1961}%
  \BibitemOpen
  \bibfield  {author} {\bibinfo {author} {\bibfnamefont {L.}~\bibnamefont
  {Faddeev}},\ }\href@noop {} {\bibfield  {journal} {\bibinfo  {journal} {Sov.
  Phys. JETP}\ }\textbf {\bibinfo {volume} {12}},\ \bibinfo {pages} {1014}
  (\bibinfo {year} {1961})}\BibitemShut {NoStop}%
\bibitem [{\citenamefont {Ethofer}\ and\ \citenamefont
  {Schuck}(1969)}]{ethofer_six-point_1969}%
  \BibitemOpen
  \bibfield  {author} {\bibinfo {author} {\bibfnamefont {S.}~\bibnamefont
  {Ethofer}}\ and\ \bibinfo {author} {\bibfnamefont {P.}~\bibnamefont
  {Schuck}},\ }\href@noop {} {\bibfield  {journal} {\bibinfo  {journal}
  {Zeitschrift f\"{u}r Physik}\ }\textbf {\bibinfo {volume} {228}},\ \bibinfo
  {pages} {264} (\bibinfo {year} {1969})}\BibitemShut {NoStop}%
\bibitem [{\citenamefont {Potthoff}\ \emph {et~al.}(1994)\citenamefont
  {Potthoff}, \citenamefont {Braun},\ and\ \citenamefont
  {Borstel}}]{potthoff_three-particle_1994}%
  \BibitemOpen
  \bibfield  {author} {\bibinfo {author} {\bibfnamefont {M.}~\bibnamefont
  {Potthoff}}, \bibinfo {author} {\bibfnamefont {J.}~\bibnamefont {Braun}}, \
  and\ \bibinfo {author} {\bibfnamefont {G.}~\bibnamefont {Borstel}},\
  }\href@noop {} {\bibfield  {journal} {\bibinfo  {journal} {Z. Phys. B}\
  }\textbf {\bibinfo {volume} {95}},\ \bibinfo {pages} {207} (\bibinfo {year}
  {1994})}\BibitemShut {NoStop}%
\bibitem [{\citenamefont {Barbieri}\ \emph {et~al.}(2007)\citenamefont
  {Barbieri}, \citenamefont {Van~Neck},\ and\ \citenamefont
  {Dickhoff}}]{barbieri_quasiparticles_2007}%
  \BibitemOpen
  \bibfield  {author} {\bibinfo {author} {\bibfnamefont {C.}~\bibnamefont
  {Barbieri}}, \bibinfo {author} {\bibfnamefont {D.}~\bibnamefont {Van~Neck}},
  \ and\ \bibinfo {author} {\bibfnamefont {W.~H.}\ \bibnamefont {Dickhoff}},\
  }\href@noop {} {\bibfield  {journal} {\bibinfo  {journal} {Phys. Rev. A}\
  }\textbf {\bibinfo {volume} {76}},\ \bibinfo {pages} {052503} (\bibinfo
  {year} {2007})}\BibitemShut {NoStop}%
\bibitem [{\citenamefont {Degroote}\ \emph {et~al.}(2011)\citenamefont
  {Degroote}, \citenamefont {Van~Neck},\ and\ \citenamefont
  {Barbieri}}]{degroote_faddeev_2011}%
  \BibitemOpen
  \bibfield  {author} {\bibinfo {author} {\bibfnamefont {M.}~\bibnamefont
  {Degroote}}, \bibinfo {author} {\bibfnamefont {D.}~\bibnamefont {Van~Neck}},
  \ and\ \bibinfo {author} {\bibfnamefont {C.}~\bibnamefont {Barbieri}},\
  }\href@noop {} {\bibfield  {journal} {\bibinfo  {journal} {Phys. Rev. A}\
  }\textbf {\bibinfo {volume} {83}},\ \bibinfo {pages} {042517} (\bibinfo
  {year} {2011})}\BibitemShut {NoStop}%
\bibitem [{\citenamefont {Schl\"{u}nzen}\ \emph {et~al.}(2020)\citenamefont
  {Schl\"{u}nzen}, \citenamefont {Joost},\ and\ \citenamefont
  {Bonitz}}]{schlunzen_achieving_2020}%
  \BibitemOpen
  \bibfield  {author} {\bibinfo {author} {\bibfnamefont {N.}~\bibnamefont
  {Schl\"{u}nzen}}, \bibinfo {author} {\bibfnamefont {J.-P.}\ \bibnamefont
  {Joost}}, \ and\ \bibinfo {author} {\bibfnamefont {M.}~\bibnamefont
  {Bonitz}},\ }\href@noop {} {\bibfield  {journal} {\bibinfo  {journal} {Phys.
  Rev. Lett.}\ }\textbf {\bibinfo {volume} {124}},\ \bibinfo {pages} {076601}
  (\bibinfo {year} {2020})}\BibitemShut {NoStop}%
\bibitem [{\citenamefont {Myhre}\ \emph {et~al.}(2019)\citenamefont {Myhre},
  \citenamefont {Coriani},\ and\ \citenamefont {Koch}}]{myhre_x-ray_2019}%
  \BibitemOpen
  \bibfield  {author} {\bibinfo {author} {\bibfnamefont {R.~H.}\ \bibnamefont
  {Myhre}}, \bibinfo {author} {\bibfnamefont {S.}~\bibnamefont {Coriani}}, \
  and\ \bibinfo {author} {\bibfnamefont {H.}~\bibnamefont {Koch}},\ }\href@noop
  {} {\bibfield  {journal} {\bibinfo  {journal} {J. Phys. Chem. A}\ }\textbf
  {\bibinfo {volume} {123}},\ \bibinfo {pages} {9701} (\bibinfo {year}
  {2019})}\BibitemShut {NoStop}%
\bibitem [{\citenamefont {Kuleff}\ and\ \citenamefont
  {Cederbaum}(2007)}]{kuleff_charge_2007}%
  \BibitemOpen
  \bibfield  {author} {\bibinfo {author} {\bibfnamefont {A.~I.}\ \bibnamefont
  {Kuleff}}\ and\ \bibinfo {author} {\bibfnamefont {L.~S.}\ \bibnamefont
  {Cederbaum}},\ }\href@noop {} {\bibfield  {journal} {\bibinfo  {journal}
  {Chem. Phys.}\ }\textbf {\bibinfo {volume} {338}},\ \bibinfo {pages} {320}
  (\bibinfo {year} {2007})}\BibitemShut {NoStop}%
\bibitem [{\citenamefont {Perfetto}\ \emph {et~al.}(2019)\citenamefont
  {Perfetto}, \citenamefont {Sangalli}, \citenamefont {Palummo}, \citenamefont
  {Marini},\ and\ \citenamefont {Stefanucci}}]{perfetto_first-principles_2019}%
  \BibitemOpen
  \bibfield  {author} {\bibinfo {author} {\bibfnamefont {E.}~\bibnamefont
  {Perfetto}}, \bibinfo {author} {\bibfnamefont {D.}~\bibnamefont {Sangalli}},
  \bibinfo {author} {\bibfnamefont {M.}~\bibnamefont {Palummo}}, \bibinfo
  {author} {\bibfnamefont {A.}~\bibnamefont {Marini}}, \ and\ \bibinfo {author}
  {\bibfnamefont {G.}~\bibnamefont {Stefanucci}},\ }\href@noop {} {\bibfield
  {journal} {\bibinfo  {journal} {J. Chem. Theory Comput.}\ }\textbf {\bibinfo
  {volume} {15}},\ \bibinfo {pages} {4526} (\bibinfo {year}
  {2019})}\BibitemShut {NoStop}%
\bibitem [{\citenamefont {Cooper}\ and\ \citenamefont
  {Averbukh}(2013)}]{cooper_single-photon_2013}%
  \BibitemOpen
  \bibfield  {author} {\bibinfo {author} {\bibfnamefont {B.}~\bibnamefont
  {Cooper}}\ and\ \bibinfo {author} {\bibfnamefont {V.}~\bibnamefont
  {Averbukh}},\ }\href@noop {} {\bibfield  {journal} {\bibinfo  {journal}
  {Phys. Rev. Lett.}\ }\textbf {\bibinfo {volume} {111}},\ \bibinfo {pages}
  {083004} (\bibinfo {year} {2013})}\BibitemShut {NoStop}%
\bibitem [{\citenamefont {Ayuso}\ \emph {et~al.}(2017)\citenamefont {Ayuso},
  \citenamefont {Palacios}, \citenamefont {Decleva},\ and\ \citenamefont
  {Mart\'{i}n}}]{ayuso_ultrafast_2017}%
  \BibitemOpen
  \bibfield  {author} {\bibinfo {author} {\bibfnamefont {D.}~\bibnamefont
  {Ayuso}}, \bibinfo {author} {\bibfnamefont {A.}~\bibnamefont {Palacios}},
  \bibinfo {author} {\bibfnamefont {P.}~\bibnamefont {Decleva}}, \ and\
  \bibinfo {author} {\bibfnamefont {F.}~\bibnamefont {Mart\'{i}n}},\
  }\href@noop {} {\bibfield  {journal} {\bibinfo  {journal} {Phys. Chem. Chem.
  Phys.}\ }\textbf {\bibinfo {volume} {19}},\ \bibinfo {pages} {19767}
  (\bibinfo {year} {2017})}\BibitemShut {NoStop}%
\bibitem [{\citenamefont {Perfetto}\ and\ \citenamefont
  {Stefanucci}(2018)}]{perfetto_cheers:_2018}%
  \BibitemOpen
  \bibfield  {author} {\bibinfo {author} {\bibfnamefont {E.}~\bibnamefont
  {Perfetto}}\ and\ \bibinfo {author} {\bibfnamefont {G.}~\bibnamefont
  {Stefanucci}},\ }\href@noop {} {\bibfield  {journal} {\bibinfo  {journal} {J.
  Phys. Condens. Matter}\ }\textbf {\bibinfo {volume} {30}},\ \bibinfo {pages}
  {465901} (\bibinfo {year} {2018})}\BibitemShut {NoStop}%
\bibitem [{\citenamefont {Li}\ \emph {et~al.}(2015)\citenamefont {Li},
  \citenamefont {Vendrell},\ and\ \citenamefont {Santra}}]{li_ultrafast_2015}%
  \BibitemOpen
  \bibfield  {author} {\bibinfo {author} {\bibfnamefont {Z.}~\bibnamefont
  {Li}}, \bibinfo {author} {\bibfnamefont {O.}~\bibnamefont {Vendrell}}, \ and\
  \bibinfo {author} {\bibfnamefont {R.}~\bibnamefont {Santra}},\ }\href@noop {}
  {\bibfield  {journal} {\bibinfo  {journal} {Phys. Rev. Lett.}\ }\textbf
  {\bibinfo {volume} {115}},\ \bibinfo {pages} {143002} (\bibinfo {year}
  {2015})}\BibitemShut {NoStop}%
\bibitem [{\citenamefont {Lara-Astiaso}\ \emph {et~al.}(2017)\citenamefont
  {Lara-Astiaso}, \citenamefont {Palacios}, \citenamefont {Decleva},
  \citenamefont {Tavernelli},\ and\ \citenamefont
  {Mart\'{i}n}}]{lara-astiaso_role_2017}%
  \BibitemOpen
  \bibfield  {author} {\bibinfo {author} {\bibfnamefont {M.}~\bibnamefont
  {Lara-Astiaso}}, \bibinfo {author} {\bibfnamefont {A.}~\bibnamefont
  {Palacios}}, \bibinfo {author} {\bibfnamefont {P.}~\bibnamefont {Decleva}},
  \bibinfo {author} {\bibfnamefont {I.}~\bibnamefont {Tavernelli}}, \ and\
  \bibinfo {author} {\bibfnamefont {F.}~\bibnamefont {Mart\'{i}n}},\
  }\href@noop {} {\bibfield  {journal} {\bibinfo  {journal} {Chem. Phys.
  Lett.}\ }\textbf {\bibinfo {volume} {683}},\ \bibinfo {pages} {357} (\bibinfo
  {year} {2017})}\BibitemShut {NoStop}%
\bibitem [{\citenamefont {Polyak}\ \emph {et~al.}(2018)\citenamefont {Polyak},
  \citenamefont {Jenkins}, \citenamefont {Vacher}, \citenamefont {Bouduban},
  \citenamefont {Bearpark},\ and\ \citenamefont {Robb}}]{polyak_charge_2018}%
  \BibitemOpen
  \bibfield  {author} {\bibinfo {author} {\bibfnamefont {I.}~\bibnamefont
  {Polyak}}, \bibinfo {author} {\bibfnamefont {A.~J.}\ \bibnamefont {Jenkins}},
  \bibinfo {author} {\bibfnamefont {M.}~\bibnamefont {Vacher}}, \bibinfo
  {author} {\bibfnamefont {M.~E.~F.}\ \bibnamefont {Bouduban}}, \bibinfo
  {author} {\bibfnamefont {M.~J.}\ \bibnamefont {Bearpark}}, \ and\ \bibinfo
  {author} {\bibfnamefont {M.~A.}\ \bibnamefont {Robb}},\ }\href@noop {}
  {\bibfield  {journal} {\bibinfo  {journal} {Mol. Phys.}\ }\textbf {\bibinfo
  {volume} {116}},\ \bibinfo {pages} {2474} (\bibinfo {year}
  {2018})}\BibitemShut {NoStop}%
\end{thebibliography}
%
\end{document}